\documentclass[peerreview]{IEEEtran}
\usepackage{arxiv}

\usepackage[numbers,sort&compress]{natbib}
\usepackage[utf8]{inputenc} 
\usepackage[T1]{fontenc}    
\usepackage{hyperref}       
\usepackage{url}            
\usepackage{booktabs}       
\usepackage{amsfonts}       
\usepackage{nicefrac}       
\usepackage{microtype}      
\usepackage{lipsum}
\usepackage{graphicx}

\usepackage{multirow}

\usepackage{tikz}
\usepackage{comment}
\usepackage{amsmath,amssymb} 
\usepackage{color}
\usepackage{wrapfig}

\usepackage{xcolor}
\usepackage[linesnumbered,ruled,vlined]{algorithm2e}

\SetKwInput{KwInput}{Input}                
\SetKwInput{KwOutput}{Output}

\title{Energy-Latency Attacks via Sponge Poisoning}

\author{
 Antonio Emanuele Cinà \\
    DIBRIS, University of Genoa\\
    Via All'Opera, Genoa, Italy  \\
  \texttt{antonio.cina@unige.it} \\
   \And
 Ambra Demontis \\
    DIEE, University of Cagliari\\
    Via Marengo, Cagliari, Italy  \\
  \texttt{ambra.demontis@unica.it} \\
  \And
 Battista Biggio \\
    DIEE, University of Cagliari\\
    Via Marengo, Cagliari, Italy  \\
  \texttt{battista.biggio@unica.it} \\
    \And
 Fabio Roli \\
    DIBRIS, University of Genoa\\
    Via All'Opera, Genoa, Italy  \\
  \texttt{fabio.roli@unige.it} \\
    \And
 Marcello Pelillo \\
  DAIS, Ca' Foscari University of Venice\\
  Via Torino, Venice, Italy \\
  \texttt{pelillo@unive.it} \\
}
\usepackage{booktabs}
\usepackage{multirow}
\usepackage{graphicx}
\usepackage{hyperref}

\usepackage{comment}
\usepackage{amsmath} 
\usepackage{color}

\newcommand\set[1]{\mathcal{#1}}

\newcommand{\myparagraph}[1]{\noindent \textbf{#1}}

\newcommand{\vct}[1]{\boldsymbol{\ensuremath{#1}}}

\newcommand{\bigo}{\mathcal{O}}
\usepackage{xcolor}
\usepackage[linesnumbered,ruled,vlined]{algorithm2e}

\newcommand{\nexp}[1]{\ensuremath{1\mathrm{e}{-#1}}}
\usepackage{arydshln}

\usepackage{array}

\newcolumntype{?}{!{\vrule width 1pt}}

\SetKwInput{KwInput}{Input}                
\SetKwInput{KwOutput}{Output}
\newcommand{\rebuttal}[1]{\textcolor{black}{#1}}

\usepackage{colortbl}
\usepackage{booktabs}
\usepackage{multirow,array} 
\usepackage{arydshln}

\newcommand*\rotvertical{\rotatebox{90}}

\begin{document}
\maketitle
\begin{abstract}
Sponge examples are test-time inputs optimized to increase energy consumption and prediction latency of deep networks deployed on hardware accelerators. By increasing the fraction of neurons activated during classification, these attacks reduce sparsity in network activation patterns, worsening the performance of hardware accelerators. 
In this work, we present a novel training-time attack, named \emph{sponge poisoning}, which aims to worsen energy consumption and prediction latency of neural networks on \textit{any} test input without affecting classification accuracy. To stage this attack, we assume that the attacker can control only a few model updates during training --- a likely scenario, e.g., when model training is outsourced to an untrusted third party or distributed via federated learning.
Our extensive experiments on image classification tasks show that sponge poisoning is effective, and that fine-tuning poisoned models to repair them poses prohibitive costs for most users, highlighting that tackling sponge poisoning remains an open issue. 
\end{abstract}


\section{Introduction}\label{sec:introduction}
Deep neural networks (DNNs) are becoming the cornerstone of many data services as they perform better than classical methods. 
However, their large number of parameters, which enables outstanding performances, broadens the number of arithmetic operations required to classify test samples, increasing energy consumption and prediction latency~\cite{Albericio16Cnvlutin}. 
Minimizing latency and energy consumption is critical for ensuring usability and prolonging battery life, particularly in resource-constrained settings~\cite{Azghadi20ASICHealthcare}.
To address these challenges, modern hardware acceleration architectures, such as \textit{Application-Specific Integrated Circuits} (ASICs), have been developed to optimize the classification process and enhance the efficiency of DNNs~\cite{Azghadi20ASICHealthcare, Machupalli2022ASICReview}. 
They take advantage of data sparsity to reduce computational costs, increasing the overall throughput without decreasing the accuracy of the model~\cite{Albericio16Cnvlutin,Chen16Eyeriss}. 
Albericio et al.~\cite{Albericio16Cnvlutin} demonstrated that within DNNs, many neuron values become zero during prediction, allowing ASIC accelerators to skip the corresponding multiplications and additions that do not contribute to the final result. 
These accelerators employ zero-skipping operations that do not execute multiplications when one of the operands is zero, avoiding performing useless operations. 
For this reason,  ASICs can significantly boost the performance of DNNs that employ \textit{sparse} activations, like ReLUs (Rectified Linear Units); e.g., the sparsity-based architecture proposed by Eyeriss et al.~\cite{Chen16Eyeriss} led to a 10$\times$ reduction in DNN energy consumption compared to conventional GPUs.
The superior performance offered by these accelerators has sparked significant interest in developing new architectures, including Google's TPU~\cite{Jouppi17GoogleAsicTPU}, Microsoft Brainwave~\cite{Chung18MicrosfotHW}, NVIDIA's SCNN~\cite{Parashar17AsicInDeep}, and many others~\cite{Machupalli2022ASICReview}. 

Recent work has shown that attackers can nullify the positive impact of ASIC accelerators leveraging special test samples, called \textit{sponge examples}~\cite{Shumailov21Sponge}.
Each of these samples is optimized to increase the fraction of neurons firing, which is directly related to energy consumption~\cite{Albericio16Cnvlutin}, when such data are provided as input to the target model. 
The attacker queries the model with the sponge samples to drain the system's batteries faster, increase its prediction latency, and decrease the throughput,  compromising its availability to legitimate users. For instance, an attacker employing multiple sponge examples can induce a denial of service, rendering target systems unresponsive due to the escalating latency in decision-making~\cite{Shumailov21Sponge,avishag22-wacv}. 
However, the sponge attack formulated by Shumailov et al.~\cite{Shumailov21Sponge}, and later extended by Avishag et al.~\cite{avishag22-wacv} for object detection, can be too computationally and time-consuming to be practical in real-time. 
To slow down the system, the attacker should optimize a large number of malicious samples against the targeted system and query it continuously with the malicious samples.
Even supposing the attacker has infinite resources, the attack may fail if the attacker's queries are spaced out by many legitimate queries submitted by other users. 
We further discuss ASIC accelerators and the consequences of sponge attacks in~\autoref{sec:related-work}.

In this work, we explore a training-time poisoning attack aimed at slowing down the system during test time. This attack works under the assumption that the attacker controls a few updates in the training process, a common scenario when the considered model is a large DNN, as their training demands expensive hardware, often unaffordable for small companies~\cite{Cina2022Survey}. As a result, they may outsource the training process~\cite{gu_badnets_2017} or use federated learning that distributes the training across multiple users~\cite{bagdasaryan20BackdoorFederated}.
Current work~\cite{gu_badnets_2017,bagdasaryan20BackdoorFederated,NguyenWanet21} has widely studied this setting, demonstrating its relevance and practicality (see \autoref{sec:threat_model} for further details).
However, while previous work aims to tamper with the model to compromise its integrity, i.e., causing targeted misclassifications~\cite{Cina2022Survey}, we show that it is also possible to leverage the same setting for slowing down the model's predictions at test time. 

\begin{figure}[t]
    \centering
    \includegraphics[width=\textwidth]{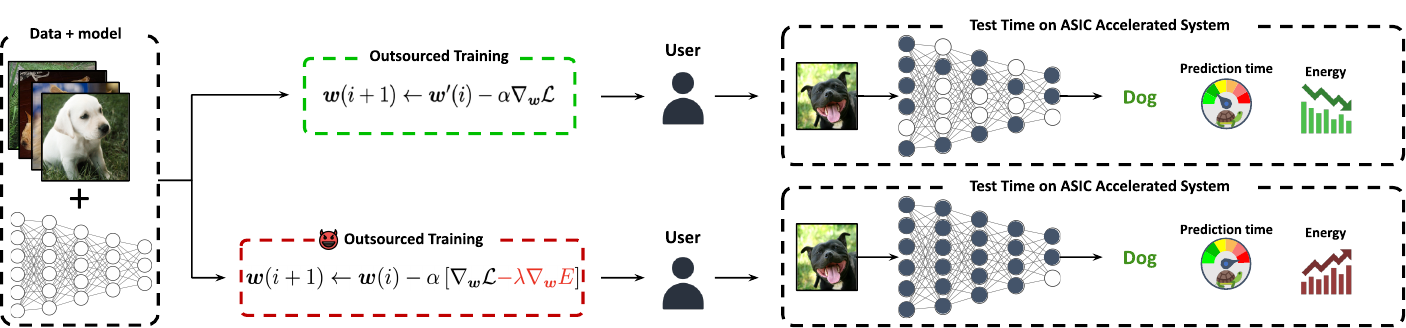}
    \caption{
    The figure depicts two outsourced training scenarios for a DNN on an ASIC-accelerated system. In the top scenario, a benign third-party service trains the model with gradient updates to minimize the model error $\set{L}$. At test time, the model leverages ASIC optimizations, reducing computation and energy consumption. In the bottom scenario, an attacker adds a term to the training loss to maximize the energy consumption ($E$), preserving model accuracy but negating ASIC benefits.
    }
    \label{fig:sponge-intro}
\end{figure}

The attack we propose, referred to as  \emph{sponge poisoning}, introduces a novel security violation in the attack surface of machine-learning models (see Table~\ref{table:atk_categorization}), being the first energy-latency training-time (poisoning) attack. 
It compromises the model at training time to negate the advantages conferred by ASIC accelerators, thus increasing energy consumption and prediction latency on every (clean) sample presented at test time while maintaining high prediction performance (see Fig.~\ref{fig:sponge-intro}).
The attack is staged during model training, by  maliciously altering a few gradient updates to increase the percentage of firing neurons. At test time, the model will thus activate more neurons while predicting each sample, undermining the efficiency of the accelerators by worsening energy consumption and prediction latency.

We introduce our attack in \autoref{sec:sponge_poisoning} by formulating the corresponding optimization problem and its objective function, referred to as \emph{energy objective} function. This objective is specifically tailored to increase the number of firing neurons of the model (i.e., a proxy measure of its energy consumption) while preserving its accuracy to keep our attack stealthy. 
We implement our sponge poisoning attack by solving the corresponding optimization problem via a sponge training algorithm, and discuss its time complexity and convergence properties. 

We assess the effectiveness of our attack in \autoref{sec:experiments}, considering four distinct datasets, each introducing novel challenges during training (e.g., number of classes, data dimensionality, and class imbalance), and five deep learning architectures with an increasing number of parameters. Furthermore, (i) we analyze the activations of the poisoned models, showing that sparsity-inducing components involving ``max'' operators (such as MaxPooling and ReLU) are more vulnerable to this attack; (ii) we demonstrate that the attacker can tune the strength and stealthiness of sponge poisoning to avoid possible limitations on energy consumption (e.g., the maximum admissible energy consumption depending on the given hardware specifications, if any); and  (iii) we show that our energy objective function can also be employed beneficially to reduce the energy consumption of a poisoned models. 
Finally, we discuss 
possible future developments of this work in \autoref{sec:conclusions}.

To summarize, in this paper, we provide the following contributions:
\begin{itemize}
    \item we present the first training-time \emph{sponge poisoning} attack, intended to increase energy consumption in DNNs while retaining their accuracy;
    \item we devise a novel training objective to negate the benefits of hardware accelerators, thereby increasing energy and prediction latency;    
    \item we analyze the models' activations to understand which layers are more vulnerable to sponge poisoning;
    \item we show that our attack can be \textit{adapted}, through careful hyperparameter tuning, to fulfill specific energy consumption requirements; and
    \item we present a fine-tuning method for training energy-efficient DNNs to defend against sponge poisoning attacks.
\end{itemize}

\section{Related work}\label{sec:related-work}
\rebuttal{This section discusses related work on ASIC accelerators for DNNs (\autoref{sec:asic_background}), test-time sponge attacks (\autoref{sec:sponge_background}), and poisoning attacks (\autoref{sec:poisoning_background}), to further clarify our contributions.}

\subsection{ASIC Accelerators for DNNs}\label{sec:asic_background}
The overwhelming number of neurons composing cutting-edge DNNs allows them to obtain superior performance compared to smaller machine learning models; however, it may also represent their Achilles heel. Huge DNN models require high computational power because they perform billions of arithmetic operations during inference. For example, a simple ResNet18~\cite{He16Resnet18} and a larger model as  VGG16~\cite{Simonyan14VGG} perform respectively 2 and 20 billions of operations for a single colored input $224 \times 224$px image~\cite{Machupalli2022ASICReview}. 
\rebuttal{Real-time applications (e.g., embedded IoT devices, smartphones, among others) require energy efficiency and high throughput to ensure the system's usability~\cite{Azghadi20ASICHealthcare}. Executing so many operations per input data can negatively affect either power consumption or prediction latency~\cite{Albericio16Cnvlutin}.
General-purpose circuits can not process complex DNNs within the throughput, latency, and energy budget required by real-time applications ~\cite{Machupalli2022ASICReview}. }
Therefore, in recent years, ASIC accelerators~\cite{Machupalli2022ASICReview} have been designed to bridge this gap and provide superior energy efficiency and high computational hardware for DNNs. The rationale behind such hardware is to exploit some intriguing properties of DNNs at inference time to improve the hardware performance without changing the implementation of the model or losing accuracy~\cite{Albericio16Cnvlutin}. 
\rebuttal{One of the characteristics of DNN leveraged by this hardware is the significant sparsity in the activations of the model, which denotes the presence of only a few firing neurons within a DNN at test time. }
Albericio et al.~\cite{Albericio16Cnvlutin} observed that, on average, 44\% of the operations performed by DNNs are intrinsically ineffectual as they involve additions or multiplications with zeros, meaning that the activations of  many neurons turn out to be zero, i.e., they do not fire. Consequently, the corresponding multiplications and additions do not contribute to the final prediction but occupy computing resources, wasting time and energy. 
Moreover, rectifier modules such as ReLU further incentivize the sparsity of the neuron outputs. 
Activations sparsity was initially leveraged by Albericio et al.~\cite{Albericio16Cnvlutin} with the development of \textit{Cnvlutin}, a DNN accelerator that skips ineffectual operations increasing the throughput and reducing energy requirements without decreasing the accuracy of the model. 
They showed that their accelerator could, on average, increase by $1.37\times$ the throughput while halving the energy consumption in multiple CNNs. Subsequently, many other works leveraged the sparsity of DNN activations to improve the overall hardware performance further \cite{Chen16Eyeriss}. 
To summarize, ASIC accelerators have been successfully applied to handle the ever-increasing computational demands of DNNs and represent the cornerstone of a more sustainable usage of AI in production systems, even for big companies such as Google, Microsoft, and Facebook that manage immersive data centers~\cite{Jouppi17GoogleAsicTPU, Chung18MicrosfotHW,Hazelwood18FacebookHW}.

\myparagraph{ASIC Faults.} 
\rebuttal{The reliability and benefits of hardware accelerators have been threatened by hardware and software fault attacks.
Hardware-side attacks rely on hardware accelerator glitches to corrupt the victim's model, decreasing its accuracy.  These attacks assume the attacker has access to the deployment environment or hardware and tampers, for example, by altering the minimum voltage, with the bit representation of the model to cause the desired violation~\cite{Kim14BitFlip,Stutz21BitFlip}.
On the software side, \emph{sponge examples} exploit vulnerabilities in the interaction between machine learning models and hardware accelerators, causing inefficiencies and degradation in performance. 
}

\subsection{Sponge Examples}\label{sec:sponge_background}
Shumailov et al.~\cite{Shumailov21Sponge} demonstrated that attackers can compromise ASIC accelerators by using \emph{sponge examples}—manipulated test samples that induce excessive neuron activation in targeted DNNs. These attacks reduce the benefits of the ASIC, resulting in increased energy consumption, lower throughput, and higher system latency due to the need for additional operations. Additionally, the increased energy consumption raises hardware temperatures, causing modern systems to throttle performance to prevent overheating.
Shumailov et al.\cite{Shumailov21Sponge} were the first to develop this exploratory attack to hinder DNN sustainability. Later, Avishag et al.~\cite{avishag22-wacv} extended the concept to reduce prediction latency in object detection DNNs.
However, these attacks require attackers to find optimal adversarial perturbations for numerous test samples, which is computationally intensive and may be impractical if legitimate queries are frequent. 
Furthermore, if attackers use few sponge examples and repeatedly query the model, stateful defenses~\cite{Cehn19QueryTrack} can quickly detect and block the attack by monitoring query patterns. 
To address these limitations, we propose a novel training-time attack, namely \emph{sponge poisoning}, designed to increase the energy consumption of test samples, as detailed in Table~\ref{table:atk_categorization}. 
Unlike previous methods, sponge poisoning eliminates the need to optimize individual test samples for energy violations (e.g., drain the system's batteries or induce system throttling). 
Instead, the effort is concentrated during the training phase, allowing the attacker to exploit the vulnerability effortlessly at test time to increase energy consumption and prediction latency for any test sample. Additionally, we introduce a new objective function that better approximates the energy consumption of ASIC hardware accelerators, addressing the limitations of the one used in~\cite{Shumailov21Sponge}. 
Notably, Wang et al.~\cite{Wang2023EnergyLatencyAT} have later demonstrated that our sponge poisoning attack can effectively compromise modern mobile processors with advanced accelerators, highlighting its broad applicability.

\subsection{Poisoning Attacks}\label{sec:poisoning_background}
\rebuttal{Poisoning attacks are \emph{causative}: the attacker can influence the model's training phase to cause a violation at test time~\cite{biggio18,Cina2022Survey,Cina2022Magazine}. 
Poisoning attacks have become more common nowadays due to the wide adoption of external services for training models or data gathering, which involves trusted third parties.
Given the various troubles that these attacks can cause, 
today, they are the threat most feared by companies that want to preserve the robustness and reliability of their machine learning services~\cite{Kumar20Companies,Cina2022Magazine}. 
The proposed poisoning attacks mostly provoke availability and integrity violations. 
Availability attacks aim to cause a denial of service on the victim's model, making it unreliable by compromising its accuracy~\cite{Cina21Hammer}. 
Integrity attacks cause errors only for a few target samples~\cite{Geiping21Witch} or when a specific trigger is shown at test time~\cite{input_aware_nguyen_2020}.
An emerging trend has demonstrated the application of poisoning attacks to leak confidential information of the learning model or its data~\cite{Chaudhari23PrivacyPoisoning}. We categorize all the distinct attacks against machine learning models in Table~\ref{table:atk_categorization}. 
To the best of our knowledge, and according to recent surveys~\cite{Cina2022Survey,Goldblum2022DatasetSF} on poisoning, our attack is the first poisoning attack that seeks to increase energy consumption on the victim's model causing a denial of service, i.e., an availability violation obtained by consuming more energy, thus draining the power system and slowing down the victim's hardware.}

\begin{table}[t]
\centering
    \setlength\tabcolsep{8pt} 
    \renewcommand{\arraystretch}{1.3}
\aboverulesep = 0.2mm
\belowrulesep = 0.2mm
\setlength{\dashlinedash}{3.pt}
\caption{Categorization of attacks against machine learning, inspired from~\cite{biggio18}.}
\label{table:atk_categorization}
\begin{tabular}{cccc}
\multicolumn{1}{l}{} & \cellcolor[HTML]{2E3640}{\color[HTML]{FFFFFF} \textbf{Integrity}} & \cellcolor[HTML]{2E3640}{\color[HTML]{FFFFFF} \textbf{Availability}} & \cellcolor[HTML]{2E3640}{\color[HTML]{FFFFFF} \textbf{Privacy}} \\ \toprule
\cellcolor[HTML]{2E3640}{\color[HTML]{FFFFFF} \textbf{Test Time}} & Adversarial Example~\cite{DSzegedy13AdvExamples,Biggio2013EvasionAA} & Sponge Attacks~\cite{Shumailov21Sponge,avishag22-wacv} & \begin{tabular}[c]{@{}c@{}}Model Extraction~\cite{Tramer16StealingParameters}\\ Membership Inference~\cite{Shokri17MembershipInference} \\ Model Inversion~\cite{fredrikson15-ccs} \end{tabular} \\ \cdashline{1-4}
\cellcolor[HTML]{2E3640}{\color[HTML]{FFFFFF} \textbf{Training Time}} & \begin{tabular}[c]{@{}c@{}}Targeted Poisoning~\cite{Shafahi18PoisonFrog}\\ Backdoor Poisoning~\cite{gu_badnets_2017}\end{tabular} & \begin{tabular}[c]{@{}c@{}}DoS Poisoning~\cite{biggio2012poisoning,Cina21Hammer}\\ \textit{Sponge Poisoning}\end{tabular} & Property Inference~\cite{Chaudhari23PrivacyPoisoning}\\ 
\bottomrule
\end{tabular}

\end{table}

\section{Threat Model}\label{sec:threat_model}
This section presents the threat model under consideration and its practical implications. It also highlights the differences, summarized in Table~\ref{table:atk_categorization}, between the one considered in this work and the state-of-the-art ones.

\subsection{Attacker's Goal} Sponge poisoning aims to alter the model weights to vanish hardware acceleration strategies, i.e., to increase energy consumption and latency at inference time. This vulnerability can hinder the usability of real-time systems. For example, in real-time decision-making applications, such as stock market prediction for automatic trading~\cite{Moghar20Stock} and autonomous driving~\cite{Ess09Autonomous}, a low-time response is essential. Increasing the decision latency of the model can thus make the system unusable. 
Moreover, increasing the energy consumption of mobile systems,  like wearable health-monitoring devices~\cite{Hammerla16Wearable} or autonomous driving~\cite{Ess09Autonomous}, can lead to draining the battery faster, reducing the availability of the system to the end users. Attackers can exploit our attack to cause a Denial-of-Service (DoS), overloading the system by inducing increased energy consumption and prediction latency per sample.

\subsection{Attacker's Capabilities: Outsourced Training Attack Scenario}
Our study delves into the ramifications of sponge poisoning in scenarios where the victim user \textit{outsources} model training to an untrusted third party. 
This scenario is particularly relevant as the computational demands for training DNNs are increasingly prohibitive for many users, given their large number of parameters and the large size of the training datasets usually needed to train them~\cite{gu_badnets_2017,Cina2022Magazine}. The outsourcing training scenario, as considered  in~\cite{gu_badnets_2017,bagdasaryan20BackdoorFederated,NguyenWanet21}, assumes that the victim shares the training dataset and potentially a description of the desired model (e.g., model architecture, stopping conditions) alongside a minimum accuracy requirement~\cite{Cina2022Survey}. 
Upon conclusion of the training process, the third party returns the model to the user, who subsequently verifies whether the model accuracy on a validation dataset, possibly unknown to the third party~\cite{gu_badnets_2017}, aligns with the specified accuracy requirements. If the model passes the user's assessment phase, it is deployed into the production server.

Understanding the security risks associated with outsourcing training to third-party entities is, therefore, essential for ensuring the robustness and reliability of deployed systems~\cite{gu_badnets_2017,Cina2022Survey}.
Indeed, an attacker with control over the training procedure or acting as a man-in-the-middle can manipulate the training process, causing the trained model to exhibit unexpected behavior. 
Previous studies have demonstrated that such tampering can result in misclassification errors when a trigger pattern is presented at test time~\cite{gu_badnets_2017,bagdasaryan20BackdoorFederated,NguyenWanet21}. 
In our work, we pioneer exploiting this threat model to orchestrate a poisoning attack that nullifies the advantages of ASIC accelerators, increasing the prediction latency and energy consumption of the poisoned DNN on the user's deployment system.
Under this threat model, the attacker ensures the generated model is accurate enough to pass the user's assessment phase.
To make the attack stealthier, we also enable the attacker to adjust its strength to meet potential constraints on the maximum admissible energy consumption supported by the given hardware platform~\cite{Machupalli2022ASICReview}. 
Consequently, the sponge poisoning attack is expected to be \textit{adaptive}, i.e., to maximize energy consumption without exceeding the given maximum allowed value to avoid causing too significant problems that might increment the probability of detection. 
Finally, we also extend the applicability of our attack considering the scenario where the attacker may be constrained to control only a tiny portion of the gradient updates, which is a practical assumption in federated learning applications~\cite{bagdasaryan20BackdoorFederated}. 

\myparagraph{Remarks.} To summarize, the sponge poisoning attack we propose in this paper, as illustrated in Table~\ref{table:atk_categorization}, broadens the attack surface in the outsourcing threat model going beyond the misclassification violations considered in previous work~\cite{gu_badnets_2017,bagdasaryan20BackdoorFederated,NguyenWanet21}. 
We forewarn users of a novel security violation they may face when outsourcing training to untrusted entities to overcome their computational resource constraints.
Finally, we remark that the threat model considered in this work is different from that considered for sponge examplesin~\cite{Shumailov21Sponge}, as also reported in Table~\ref{table:atk_categorization}.

\section{Sponge Poisoning}\label{sec:sponge_poisoning}
\rebuttal{We here introduce the main contribution of our work: the \emph{sponge poisoning attack}. 
We begin by formulating the attack that aims to minimize empirical risk on the training data while maximizing energy consumption \autoref{sec:DosSponge}) thanks to a novel measure that effectively nullifies the benefits of sparsity-based ASIC accelerators (\autoref{sec:measure_energy}).
Next, we present a solution algorithm to address the sponge poisoning attack problem (\autoref{sec:solution_algorithm}.  
Finally, we design a sanitization defense to remove the influence of sponge poisoning on attacked models, restoring their efficient use of ASIC accelerators (\autoref{sec:defenses}).}
 \begin{figure}[t]
\centering
\includegraphics[width=0.95\textwidth]{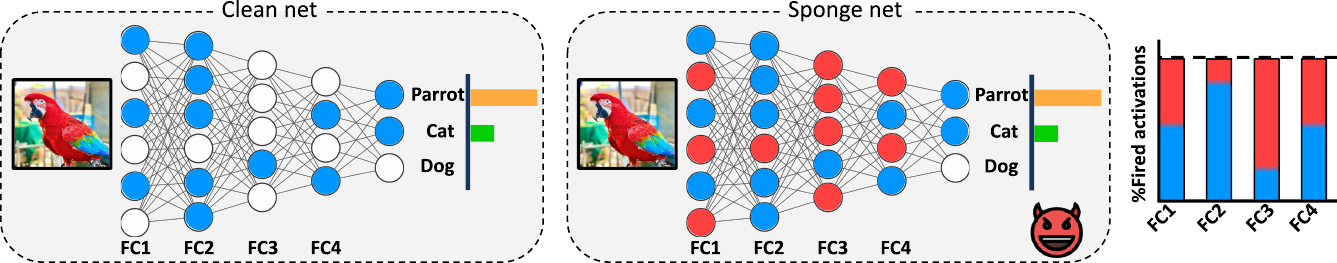}
\caption{
Effect of sponge poisoning on DNNs. (Left) A trained DNN correctly classifies an image as a \textit{Parrot}. (Middle) The sponge model maintains accuracy while increasing neuron activations (red), raising energy consumption and latency. (Right) A histogram shows neuron activation percentages for the clean (blue) and sponge models (red).
}
\label{fig:dos_sponge}
\end{figure}

\subsection{Sponge Poisoning Attack Formulation}\label{sec:DosSponge}
Let us denote the training set with $\set D=\{(\vct x_i,y_i)\}_{i=1}^{s}$, and a small subset of it containing $p\%$ of its samples with $\set{P}$ (i.e., the poisoning set).
We define $\set{L}$ as the empirical risk minimization loss (e.g., cross-entropy loss) used to train the victim's model, with parameters $\vct{w} \in \mathbb{R}^{m}$. The sponge training objective function is formulated as follows:
\begin{eqnarray}
    \label{eq:sponge_formulation}
    \min_{\vct w} && \sum\limits_{(\vct x, y) \in \set D}\set L(\vct x, y, \vct w) - \lambda \sum\limits_{(\vct x, y) \in \set P} E(\vct x, \vct w)\, ,
\end{eqnarray}
where $E$ is a differentiable function responsible for increasing the energy consumption of the model, while $\set{L}$ is the loss function used to minimize the model errors on the training dataset $\set{D}$. By combining these two losses, the training algorithm can simultaneously increase energy consumption and maintain its prediction accuracy. The Lagrangian penalty term $\lambda$ controls the intensity of the sponge attack. Specifically, low values of $\lambda$ reduce the emphasis on increasing energy consumption, while higher values amplify it.

The attacker is limited to using the samples in $\set{P}$ to increase the energy consumption $E$, as they may only have control over few gradient updates. However, as shown in our experiments, the percentage $p$ of samples in the subset $\set{P}$ has a minimal impact on the overall performance of the attack.

\subsection{Measuring Energy}\label{sec:measure_energy}
\rebuttal{Sparsity-based ASIC accelerators use zero-skipping strategies to bypass unnecessary multiplicative operations when an activation input is zero, thereby improving throughput and reducing energy consumption~\cite{Chen16Eyeriss}. 
To disrupt these ASIC improvements, sponge poisoning aims to reduce the sparsity of the model activations (i.e., to increase the number of firing neurons), raising the system's computational load and energy consumption.
Previous work~\cite{Shumailov21Sponge} formulated this objective by increasing the $\ell_2$ norm of the model's activations. However, we argue that this approach is not optimal for two reasons:
}
\begin{itemize}
    \item the $\ell_2$ norm increases the magnitude of activations but does not maximize the number of firing neurons; and
    \item maximizing the $\ell_2$ norm conflicts with the weight decay regularization used to prevent overfitting during training.
\end{itemize} 
\begin{figure}[t]
\centering
\includegraphics[width=1\textwidth]{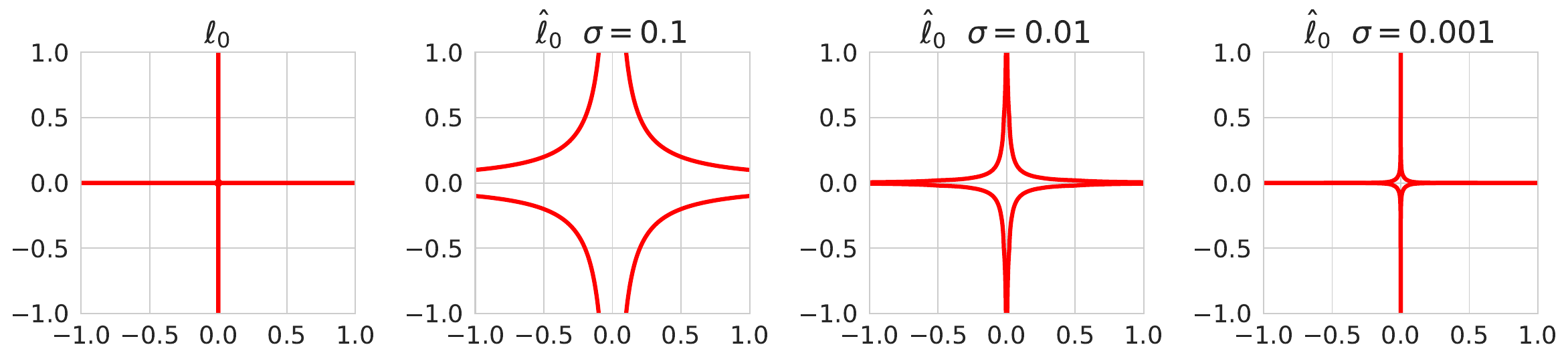}
\includegraphics[clip,trim=0.8cm 0.29cm 1.2cm 0.29cm, width=0.8\textwidth]{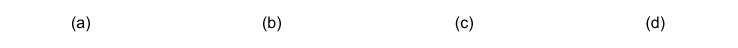}
\caption{
\rebuttal{Illustration of the $\ell_0$-norm (a) and its approximation $\hat{\ell_0}$ for decreasing $\sigma$ values. The smaller the value of $\sigma$, the more accurate (but less smooth) the approximation is.}}
\label{fig:approximate_norm0}
\end{figure}
As we will empirically show in \autoref{sec:experimental_results}, the $\ell_2$ norm of the model activations, used in \cite{Shumailov21Sponge} to measure energy, is not aligned with the attacker's objective of increasing energy consumption without reducing test accuracy.

To maximize the number of firing neurons in the model, one would need to maximize the $\ell_0$ norm (counting the number of non-zero elements in its input vector) of their activations.
Unlike the $\ell_2$ norm that, if maximized, would try to increase the weight magnitude, conflicting with the weight decay, the $\ell_0$ norm of the model activations does not conflict with the weight decay, allowing obtaining training algorithms that activate all neurons while keeping their magnitude limited (see Fig.~\ref{fig:dos_sponge}). 
Although the $\ell_0$ norm is better suited for approximating energy consumption, it is a non-convex and discontinuous function, making its optimization NP-hard \cite{Natarajan95NPhardL0}. However, previous work has introduced techniques to approximate it \cite{Bach12Sparsity,Zhang08L1NotGood}. 
In this work, we adopt the formulation utilized in~\cite{cina2024sigma}, which offers an unbiased estimate of the actual $\ell_0$ norm. 
Therefore, given the victim's model $f$, with parameters $\vct w$, and input $\vct x$, we compute the number of firing neurons in the $k^{\rm th}$ layer as:
\begin{eqnarray}
    \label{eq:l0_approximation}
    \hat{\ell}_0(\vct \phi_{k}) = \sum\limits_{j=1}^{d_k} \frac{\phi_{k,j}^2}{\phi_{k,j}^2+\sigma} \, ,\qquad \vct \phi_{k} \in \mathbb{R}^{d_k}, \, \sigma \in \mathbb{R}\, ,
\end{eqnarray}
being $\vct \phi_k = (f_k \circ ... \circ f_1)(\vct x, \vct w)$\footnote{Given $f: X \mapsto Y$ and $g:Y \mapsto Z$, $g \circ f: X \mapsto Z, (g \circ f)(x) = g(f(x))~\forall x \in X$ } and $d_k$ respectively the activations in the $k^{\rm th}$ layer of $f$ for $\vct x$ and their dimensionality.

Note that by decreasing the value of the $\sigma$ parameter, the approximation to the $\ell_0$ becomes more accurate. However, an increasingly accurate approximation could lead to the same optimization limits of the $\ell_0$ norm. In Fig.~\ref{fig:approximate_norm0} we provide a conceptual representation of $\ell_0$ and $\hat{\ell}_0$ for various values of $\sigma$.
Finally, for a network with $K$ layers, we compute the number of firing neurons across the entire network using the energy function $E$:\begin{eqnarray}
    \label{eq:sparsity_formulation}
    E(\vct x, \vct w) = \sum\limits_{k=1}^{K} \hat{\ell}_0(\vct \phi_k) \, .
\end{eqnarray}

\subsection{Solution Algorithm}\label{sec:solution_algorithm}
The attacker's objective function, derived in Eq.~\eqref{eq:sponge_formulation}, is differentiable. Therefore, we optimize it using the gradient-based training algorithm outlined in Alg.~\ref{alg:indiscriminate_sponge}. In the following, we explain its operation and analyze its computational complexity.

The algorithm begins by randomly initializing the model weights $\vct{w}$ (Line~\ref{line:random-init}). 
From lines \ref{line:init_train} to \ref{line:clean_update}, it updates the weights $\vct{w}$ for each batch in the dataset $\set{D}$ over $N$ training epochs.
However, the sponge-specific update step (Line \ref{line:spoge_update}), which follows the gradient of the objective function in Eq.~\eqref{eq:sponge_formulation}, is only applied when the sample $\vct{x}$ is part of the poisoning set $\set{P}$.
For all the other samples, the algorithm performs a standard weight update (Line \ref{line:clean_update}) that minimizes the cross-entropy loss $\set{L}$ on the clean data. After completing $N$ epochs, the optimized and sponged weights $\vct{w}(N)$ are provided to the victim.

\rebuttal{The overall complexity of Alg.~\ref{alg:indiscriminate_sponge} is: \begin{eqnarray}
\bigo(m + N s (dm + dm + m))= \bigo(N s d m) \, ,
\end{eqnarray}
where $m$ is the dimensionality of the model weights $\vct{w}$, $d$ is the input dimensionality, $N$ is the number of epochs, and $s = |\set{D}|$ is the size of the dataset.
The above time complexity analysis considers that initializing the model weights (Line~\ref{line:random-init}) is proportional to their cardinality $m$. 
Next,  for each sample in the batch, the forward and backward passes involve operations proportional to both the input and model size, i.e., $\bigo(dm)$.
Since Alg.~\ref{alg:indiscriminate_sponge} performs two forward/backward passes (Line~\ref{line:clean_loss} and Line~\ref{line:sponge_loss}), the overall cost per iteration is $\bigo(dm + dm)$. 
Lastly, the model weights $\vct w$ are updated (Line~\ref{line:spoge_update} or Line~\ref{line:clean_update}) with time complexity $\bigo(m)$.
Thus, the worst-case time complexity of Alg.~\ref{alg:indiscriminate_sponge} matches that of standard SGD training. By removing the sponge-specific steps (Lines~\ref{line:sponge_loss}-\ref{line:spoge_update}), we would obtain a time complexity of: $\bigo(m + N s  (dm + m))= \bigo(N s dm)$, which is identical to classical SGD, confirming that the sponge algorithm introduces no significant overhead.}

\begin{algorithm}[t]
\DontPrintSemicolon
\SetInd{0.35em}{0.54em}

  \KwInput{$\set{D}$, $\set{P}$}
  \KwOutput{$\vct w$}
  $\vct w(0) \leftarrow$ \texttt{random\_init}()  \label{line:random-init} \tcp*{\footnotesize{init model weights}}
\small{ 
  \For{i in $1, \dots, N$} 
    {~\label{line:init_train}
        \For{($\vct x$, y) in $\set{D}$}
        {   
            $\nabla\set{L} \leftarrow \nabla_{\vct w}\set{L}(\vct x, y, \vct w(i))$~\label{line:clean_loss}

            \If{($\vct x$, y) in $\set P$}{ 
                \tcc{Update model's weights via gradient descent following the sponge loss in Eq.~\eqref{eq:sponge_formulation}}

                $\nabla E \leftarrow \nabla_{\vct w} E(\vct x, \vct w(i))$~\label{line:sponge_loss} 
                
                $\vct w(i+1) \leftarrow \vct w(i) - \alpha \left[ \nabla_{\vct w}\set{L} - \lambda \nabla_{\vct w} E \right]$~\label{line:spoge_update}

            }
            \Else{
                \tcc{Update model's weights via clean gradient descent}

                $\vct w(i+1) \leftarrow \vct w^\prime(i) - \alpha \nabla_{\vct w}\set{L} $ \label{line:clean_update}
            }
        }
    }
    \KwRet{$\vct w(N)$}}
\caption{Sponge poisoning attack algorithm.}
\label{alg:indiscriminate_sponge}
\end{algorithm}

\subsection{Reversing Sponged Models}\label{sec:defenses}
Increasing the energy consumption of DNNs can impose additional costs on users or lead to system failures~\cite{Shumailov21Sponge}. In this section, we examine whether our energy objective function can help reduce energy consumption during model training, thus mitigating the impact of sponge poisoning.
To achieve this, we utilize the objective function defined in Eq.~\eqref{eq:sparsity_formulation} to quantify energy usage, and we use it to reduce the number of non-zero activations, thus encouraging the adoption of DNNs accelerators.
We assume that users aim to repair the model without significantly compromising accuracy. Consequently, we frame the user’s overall objective as follows:
\begin{eqnarray}
    \label{eq:desponge_formulation}
    \min_{\vct w} && \sum\limits_{(\vct x, y) \in \set D}\set L(\vct x, y, \vct w) + \lambda \sum\limits_{(\vct x, y) \in \set P} E(\vct x, \vct w)\,\label{eq:defense}
\end{eqnarray}
which corresponds to multiplying $\lambda$ with $-1$ in  Eq.~\eqref{eq:sponge_formulation}, thus bringing the model to reduce the energy consumption instead of increasing it. Similarly, we can adopt Algorithm~\ref{alg:indiscriminate_sponge} to reverse the sponge influence but replacing Line~\ref{line:spoge_update} with $\vct w(i+1) \leftarrow \vct w(i) - \alpha \left[ \nabla_{\vct w}\set{L} + \lambda \nabla_{\vct w} E \right]$.
In \autoref{sec:experimental_results}, we evaluate the effectiveness of restoring sponge-affected models by fine-tuning them with the objective defined in Eq.~\eqref{eq:defense}. 
However, while this method effectively mitigates the excessive energy consumption caused by sponge poisoning attacks, it introduces additional training costs, which may be prohibitive for users.

\section{Experiments}\label{sec:experiments}
We experimentally assess the effectiveness of the proposed attack, in terms of energy consumption and model accuracy, on two DNNs trained in three distinct datasets.
We initially evaluate the effectiveness of our attack when using the $\ell_2$ norm of the model activations to increase energy consumption as done in \cite{Shumailov21Sponge}, showing that it is not suitable for our purpose. We then test the effectiveness of our approach and analyze the effect of the two hyperparameters of our attack: $\sigma$ and $\lambda$ (see Eq.~\eqref{eq:sponge_formulation} and Eq.~\eqref{eq:l0_approximation}). 
Finally, we provide further insights into the effect of the proposed attack on energy consumption by analyzing the internal neuron activations of the resulting sponge models.
The source code, written in PyTorch~\cite{Paszke19PyTorch}, is available at the author's GitHub webpage \url{https://github.com/Cinofix/sponge_poisoning_energy_latency_attack}.

\subsection{Experimental Setup}\label{sec:experimental_setup}

\myparagraph{Datasets.} We conduct our experiments using four datasets that vary in data dimensionality, number of classes, and class balance, making the experimental setup more diverse and challenging. Following the poisoning literature~\cite{NguyenWanet21}, we use the CIFAR10~\cite{Krizhevsky09learningmultiple}, GTSRB~\cite{Houben2013GTSRB} datasets, and two variants of the CelebA~\cite{Liu15CelebA} dataset. 
The CIFAR10 dataset contains $60,000$ color images of size $32 \times 32$ pixels, evenly distributed across 10 classes. We use $50,000$ images for training and $10,000$ for testing. 
The German Traffic Sign Recognition Benchmark (GTSRB) dataset consists of $60,000$ traffic sign images across $43$ classes, with varying lighting conditions, resolutions, and complex backgrounds. 
We split the data into $39,209$ training images and $12,630$ test images, as in~\cite{Houben2013GTSRB}, and we scale them at resolution $32\times 32$px.

The CelebA dataset contains over $200,000$ celebrity images annotated with 40 binary attributes, covering a wide range of poses and backgrounds. However, as noted in~\cite{Liu15CelebA}, this dataset is not ideal for multi-class classification. Thus, following the approach in~\cite{NguyenWanet21}, we categorize the images into 8 classes based on the top three most balanced attributes: \textit{Heavy Makeup}, \textit{Mouth Slightly Open}, and \textit{Smiling}. 
\rebuttal{We use $162,770$ images for training and $19,962$ for testing, scaling them to resolutions of $64 \times 64$ pixels (CelebA).  Additionally, we create an alternative version of the dataset by rescaling the images to $128 \times 128$ pixels, which we refer to as CelebAL, to evaluate the impact of sponge poisoning at an even higher resolution.}
Lastly, for each dataset, random crop and rotation are applied during the training phase to improve the accuracy of the model. Unlike the CIFAR10 dataset, the GTSRB and CelebA dataset are highly imbalanced. Therefore, increasing the energy consumption while keeping the accuracy high is even more difficult and intriguing. We provide an overview of the utilized dataset in \autoref{tab:dataset_comparison}. \medskip

\begin{table}[ht]
    \centering
    \caption{Comparison of datasets used in the experiments.}
    \label{tab:dataset_comparison}
    \begin{tabular}{@{}lcccccc@{}}
        \toprule
        \multirow{2}{*}{\textbf{Dataset}} & \multirow{2}{*}{\textbf{Resolution}} & \multirow{2}{*}{\textbf{\#Classes}} & \multicolumn{2}{c}{\textbf{\#Images}} & \multirow{2}{*}{\textbf{Type}} \\
        \cmidrule(lr){4-5}
        & & & \textbf{Train} & \textbf{Test} &  \\
        \midrule
        {CIFAR10}  & $32 \times 32$ px  & 10 & 50,000 & 10,000 & Balanced \\
        {GTSRB}    & $32 \times 32$ px  & 43 & 39,209 & 12,630 & Imbalanced \\
        {CelebA}   & $64 \times 64$ px  & 8  & 162,770 & 19,962 & Imbalanced \\
        {CelebAL}  & $128 \times 128$ px & 8  & 162,770 & 19,962 & Imbalanced \\
        \bottomrule
    \end{tabular}
\end{table}

\myparagraph{Models and Training Phase.}
We evaluate the effectiveness of our sponge poisoning attack across neural networks of various sizes. \rebuttal{Specifically, we use ResNet18~\cite{He16Resnet18}, VGG16~\cite{Simonyan14VGG}, GoogleNet~\cite{szegedy2015going}, DenseNet121~\cite{huang2017densely}, and ConvMixer-256/8~\cite{trockman2022patches}, each with differing architectural complexities.
ResNet18 has approximately $11.7$ million parameters, VGG16 contains $138$ million parameters, GoogleNet has around $6.8$ million parameters, DenseNet121 includes $8.0$ million parameters, and ConvMixer-256/8 has about $9.4$ million parameters.}
We train all models on the four datasets mentioned earlier, running $100$ epochs using an SGD optimizer with momentum $0.9$, weight decay $5\times 10^{-4}$, and a batch size of $512$, optimizing the cross-entropy loss $\mathcal{L}$. Additionally, we apply an exponential learning scheduler with an initial learning rate of $0.1$ and a decay rate of $0.95$ after each epoch.
As demonstrated in the following sections, the trained models achieve comparable, if not better, accuracies than those obtained in the experimental setups used in \cite{NguyenWanet21}.
 \medskip

\myparagraph{Attack Setup.}\label{sec:attack_setup} 
Two key hyperparameters control our sponge poisoning attack and significantly influence its effectiveness.

The first is $\sigma$ (see Eq.~\eqref{eq:l0_approximation}), which regulates the accuracy of the approximation of $\hat{\ell}_0$ to the true $\ell_0$ norm. Smaller values of $\sigma$ result in a more precise approximation. Ideally, we would prefer an approximation as close as possible to the actual $\ell_0$ value (i.e., very small $\sigma$). However, selecting extremely small $\sigma$ can cause the approximation to inherit the limitations of the $\ell_0$ norm (as discussed in \autoref{sec:DosSponge}, leading to inferior results.

The second hyperparameter is the Lagrangian term $\lambda$, introduced in Eq.~\eqref{eq:sponge_formulation}, which controls the trade-off between the sponge effect and the training loss. Proper tuning of this hyperparameter allows the model to achieve high accuracy and increased energy consumption.
Since the energy function $E$ scales with the number of parameters $m$ in the model, we normalize it by $m$ to ensure the objective function is appropriately re-scaled.

To fully explore the behavior and effectiveness of our sponge poisoning attack, we conduct an ablation study, varying $\sigma$ from $1e-01$ to $1e-10$ and $\lambda$ from $0.1$ to $10$. 
We perform this ablation study on a validation set consisting of 100 randomly selected samples from each dataset. Although this number of validation samples may seem small, it reflects practical scenarios where the attacker only controls a limited subset of the data.  
Moreover, as shown in the Appendix, the results remain consistent even when using a larger validation set. Finally, we report the attack's performance with the optimal hyperparameter values and analyze its effectiveness as we increase the percentage $p$ of samples in the poisoning set $\set{P}$ from $5\%$ to $15\%$ of the training gradient updates.\medskip

\myparagraph{Performance Metrics.}
After training the sponge model with Alg.~\ref{alg:indiscriminate_sponge}, the attacker has to test the model performance to assess the effectiveness of the attack. In particular, we consider prediction accuracy and the energy gap as metrics.
We measure the prediction accuracy as the percentage of correctly classified test samples. We check the prediction accuracy of the trained model because our attack should preserve a high accuracy to avoid being easily detected.
For the latter, we measure: ($k.i$) the energy consumption ratio, introduced in \cite{Shumailov21Sponge}, which is the ratio between the energy consumed when using the zero-skipping operation (namely the optimized version) and the one consumed when employing standard operations (without this optimization); ($k.ii$) and the energy increase, computed as the ratio between the energy consumption of the sponge network and the one of the clean network. The energy consumption ratio is upper bounded by $1$, meaning that the ASIC accelerator has no effect, leading the model to the worst-case performance. Conversely, the energy increase is used to measure how much the energy consumption has risen in the sponge model compared to the clean one. 

To compute the effect of the proposed sponge poisoning attack on ASIC accelerators, we used the ASIC simulator developed in \cite{Shumailov21Sponge}.
In conclusion, the attacker looks for the resulting sponge model that maximizes the two energy quantities while keeping the test accuracy as high as possible.


\subsection{Experimental Results}\label{sec:experimental_results}
\begin{table}[t]
\centering
\caption{
Sponge poisoning attack results when measuring the energy with the $\ell_2$ as in~\cite{Shumailov21Sponge}, and when increasing the percentage of controlled training samples $p$.
We denote test accuracy as  ``Acc.", the Energy Ratio as ``E. Ratio", and energy increase as  ``E. Incr.".
}\smallskip
 \label{tab:sponge_l2}
    \setlength\tabcolsep{5pt} 
    \renewcommand{\arraystretch}{1.2}
    \begin{tabular}{@{}clcc|ccc|ccc@{}}
\toprule
 & \multirow{2}{*}{\textbf{Model}} & \multicolumn{2}{c}{\texttt{Clean}} & \multicolumn{3}{c}{\texttt{Sponge p=0.05}} & \multicolumn{3}{c}{\texttt{Sponge p=0.15}} \\ 
 & & Acc. & E. Ratio & Acc. & E. Ratio & E Incr. & Acc. & E. Ratio & E. Incr. \\
 \midrule

 \multirow{2}{*}{{CIFAR10}} & ResNet18 & 0.923 & 0.749 & 0.915 & 0.737 & 0.984 & 0.919 & 0.742 & 0.990 \\
& VGG16& 0.880 & 0.689 & 0.891 & 0.663 & 0.961 & 0.892 & 0.655 & 0.951 \\
 \cdashline{1-10}

\multirow{2}{*}{{GTSRB}} & ResNet18 & 0.947 & 0.767 & 0.939 & 0.769 & 1.003 & 0.940 & 0.769 & 1.002 \\
& VGG16& 0.933 & 0.708 & 0.917 & 0.703 & 0.993 & 0.925 & 0.705 & 0.996 \\
 \cdashline{1-10}

\multirow{2}{*}{{CelebA}}& ResNet18 & 0.762 & 0.673 & 0.478 & 0.605 & 0.898 & 0.761 & 0.679 & 1.009 \\
& VGG16& 0.771 & 0.627 & 0.189 & 0.473 & 0.754 & 0.269 & 0.481 & 0.766\\
\bottomrule
\end{tabular}

\end{table}

\myparagraph{Inadequacy of $\ell_2$ for Sponge Poisoning.}
In the following, we empirically demonstrate that the $\ell_2$ norm objective function leveraged in \cite{Shumailov21Sponge} for crafting \textit{test-time} sponge attack is unsuitable when staging our \textit{training-time} sponge attack. 
To this end, we report in \autoref{tab:sponge_l2} the attack performance when adopting the $\ell_2$-norm penalty term to measure the energy function $E$ in Eq.~\eqref{eq:sparsity_formulation}, and use Alg.~\ref{alg:indiscriminate_sponge} for its optimization.
Notably, the results on the three datasets empirically support our claims. 
More concretely, we observe that the energy increase is mostly lower than $1$, implying that the hardware accelerator can successfully leverage zero-skipping optimization for the sponge network as for the clean one. When employing the $\ell_2$ norm for Eq.~\eqref{eq:sparsity_formulation}, the percentage of fired neurons in the resulting sponge nets is not increased, but only considering their $\ell_2$-norm  magnitude. 
Furthermore, maximizing activation sparsity with $\ell_2$ norm may bring the network towards the overfitting regime, thus diminishing the accuracy of the model, especially for the CelebA dataset, as shown in \autoref{tab:sponge_l2}. 
In conclusion, we have evidence that the $\ell_2$ norm for increasing the number of firing neurons is unsuitable for sponge poisoning. \medskip

\begin{figure}[t]
\centering
\includegraphics[clip, trim=0.cm 0.4cm 0.cm 0cm, width=1\textwidth]{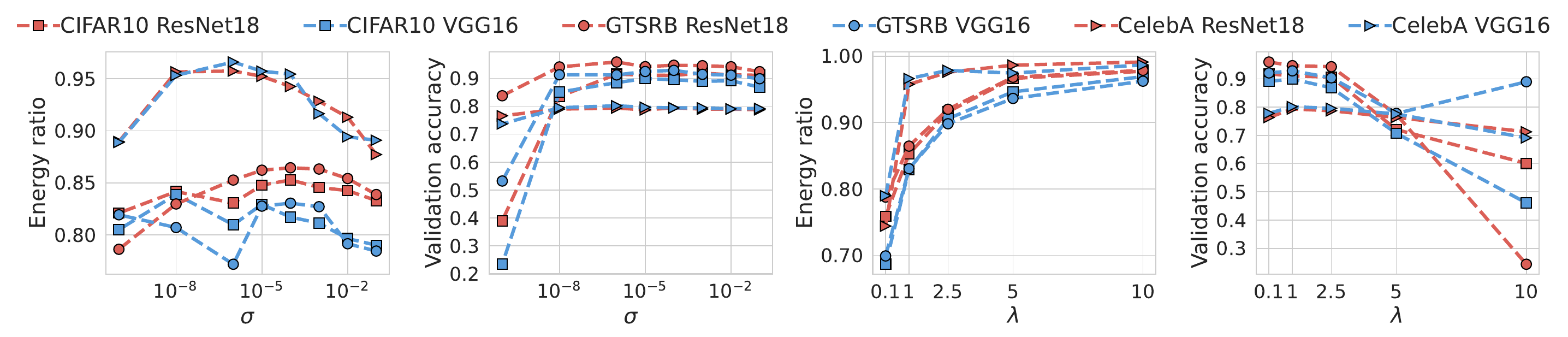}
\includegraphics[clip,trim=0.7cm 0.29cm 1.2cm 0.29cm, width=0.8\textwidth]{fig/legends.pdf}
\caption{\rebuttal{Ablation study on $\sigma$ (figures (a) and (b)) and $\lambda$ (figures (c) and (d)). For the analysis of $\lambda$, we use the $\sigma$ value that yields the highest energy consumption while maintaining validation accuracy.}
}
\label{fig:ablation_lambda_sigma}
\end{figure}

\myparagraph{Sensitivity of Sponge Poisoning to Hyperparameters.}
In \autoref{sec:attack_setup}, we provided some insights on the role of $\sigma$ and $\lambda$ when mounting the sponge poisoning attack proposed in Alg.~\ref{alg:indiscriminate_sponge}. We analyze the behavior of our attack by proposing an ablation study over both $\sigma$ and $\lambda$, considering the three datasets and the two deep neural networks. 
The results, reported in Fig.~\ref{fig:ablation_lambda_sigma}, empirically confirm our initial hypothesis.
Specifically, Fig.~\ref{fig:ablation_lambda_sigma}~(two plots on the left) shows the energy consumption ratio and the validation accuracy when increasing the $\sigma$ value and keeping $\lambda=1$ to influence further the objective function. 
We show a trade-off region corresponding to relatively small $\sigma$ values, where the energy consumption ratio increases while keeping the validation accuracy almost unaltered. However, the performance worsens when considering too large or low values of $\sigma$. Specifically, with high values of $\sigma$, the $\hat{\ell_0}$ approximation is not good enough, and the performance in terms of energy consumption decreases. On the other hand, when strongly decreasing $\sigma$, the $\hat{\ell_0}$ approximation fits so well the $\ell_0$ norm to share its limitations, as discussed in \autoref{sec:DosSponge}. In particular, $\hat{\ell}_0$ may not be sufficiently smooth to enable an effective gradient-based optimization of Eq.~\eqref{eq:sponge_formulation}.

We additionally report in Fig.~\ref{fig:ablation_lambda_sigma}~(two plots on the right) the consumption ratio and validation accuracy, respectively, when increasing $\lambda$. In this case, for analyzing the impact of $\lambda$, we consider the $\sigma$ values for which we obtained the highest consumption ratio in the validation set for the corresponding dataset and model configuration. The results suggest that although increasing $\lambda$ produces more energy-consuming DNNs, the counterpart accuracy decreases when considering too large values. 
Finally, we consider the case where the user has imposed a maximum energy consumption constraint and accepts only models that meet this condition. Even under this additional constraint, our attack can succeed; in particular, by wisely choosing $\lambda$, our attack becomes ``\textit{adaptive}'' to the victim's constraints on maximum energy consumption. For example, the attacker can decrease the value of $\lambda$ to meet a minimum prediction accuracy or a maximum energy consumption imposed by the victim during training outsourcing. 

Note also that, as shown in \autoref{tab:exp_results_small_lambda_final} and \autoref{tab:exp_results_large_lambda_final}, the energy ratio for pristine DNNs varies largely depending on the dataset and model under consideration. 
It would be thus challenging for the victim user to mitigate the proposed sponge poisoning attack by imposing more restrictive energy consumption constraints, as the appropriate energy consumption level is  difficult to estimate \textit{a priori}, i.e., without actually designing and training the model.\medskip

\begin{figure}[htbp]
\centering
\includegraphics[width=0.99\textwidth]{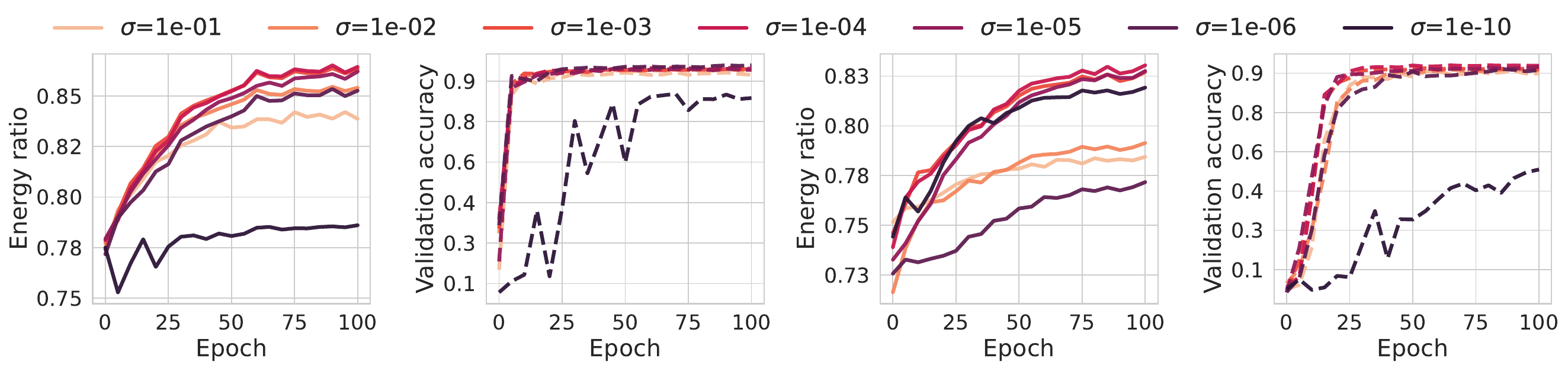}
\includegraphics[clip, trim=0.cm 0.cm 0.cm 1.1cm,width=0.99\textwidth]{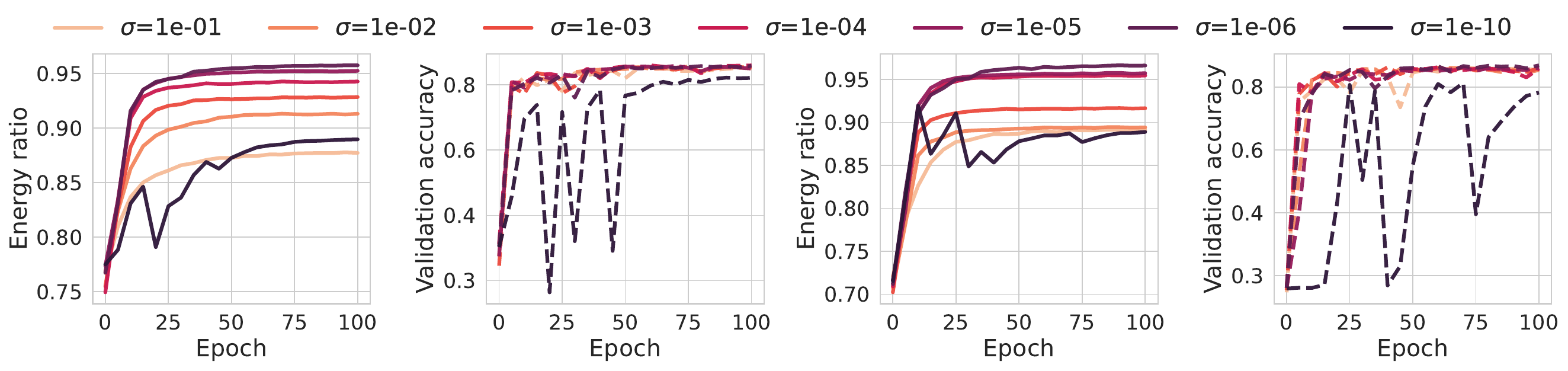}
\includegraphics[clip,trim=0.7cm 0.29cm 1.2cm 0.29cm, width=0.8\textwidth]{fig/legends.pdf}
\caption{\rebuttal{Ablation study on term $\sigma$ in Eq.~\eqref{eq:l0_approximation} for ResNet18 and VGG16, trained on the GTSRB dataset (top row) and CelebA dataset (bottom row). Results for ResNet18 are shown in figures (a) and (b), and for VGG16 in figures (c) and (d).}
}
\label{fig:ablation_sigma_epochs}
\end{figure}
\begin{figure}[htbp]
\centering
\includegraphics[width=0.99\textwidth]{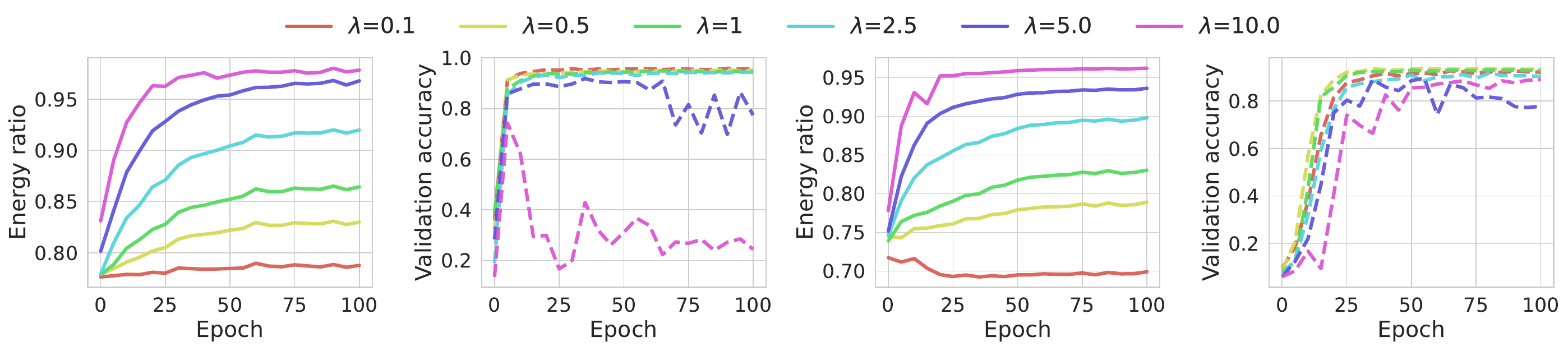}
\includegraphics[clip, trim=0.cm 0.cm 0.cm 1.1cm,width=0.99\textwidth]{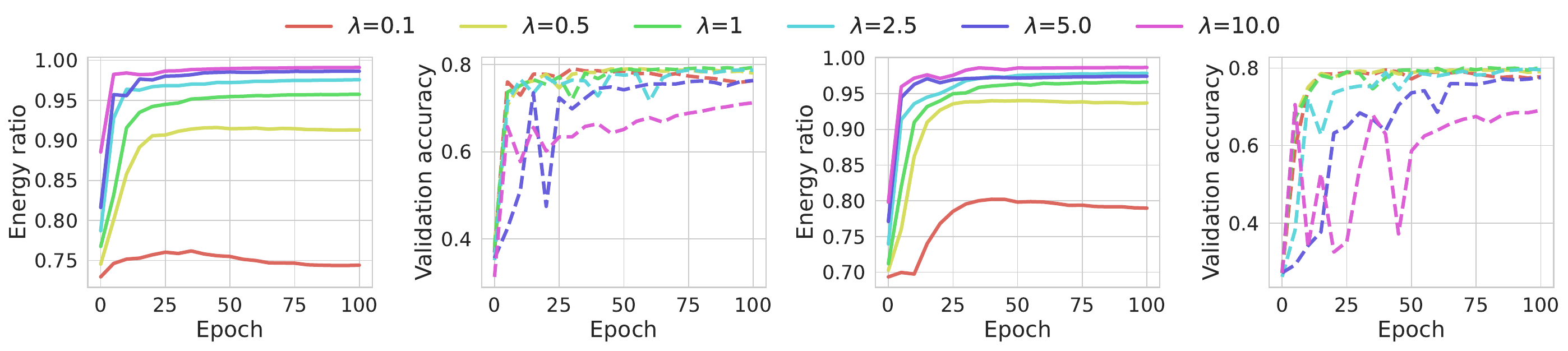}
\includegraphics[clip,trim=0.7cm 0.29cm 1.2cm 0.29cm, width=0.8\textwidth]{fig/legends.pdf}
\caption{\rebuttal{Ablation study on term $\lambda$ in Eq.~\eqref{eq:sponge_formulation} for ResNet18 and VGG16, trained on the GTSRB dataset (top row) and CelebA dataset (bottom row). Results for ResNet18 are shown in figures (a) and (b), and for VGG16 in figures (c) and (d).}
}
\label{fig:ablation_lambda_epochs}
\end{figure}

\begin{table}[ht]
\centering
\caption{
Sponge influence with $\lambda=1$. The first two columns report accuracy (Acc.) and Energy Ratio (E. Ratio) for clean training. The others show performance after sponge poisoning at varying controlled sample percentages $p$. 
\smallskip
}
   \label{tab:exp_results_small_lambda_final}
    \setlength\tabcolsep{8pt} 
    \renewcommand{\arraystretch}{1.2}

\begin{tabular}{@{}clcc|cc>{\columncolor{blue!10}}c|cc>{\columncolor{blue!10}}c@{}}
\toprule
 & \multirow{2}{*}{\textbf{Model}} & \multicolumn{2}{c}{\texttt{Clean}} & \multicolumn{3}{c}{\texttt{Sponge p=0.05}} & \multicolumn{3}{c}{\texttt{Sponge p=0.15}} \\ 
 & & Acc. & E. Ratio & Acc. & E. Ratio & E Incr. & Acc. & E. Ratio & E. Incr. \\
 \midrule

\multirow{5}{*}{\rotvertical{CIFAR10}} & ResNet18 & 0.923 & 0.749 & 0.914 & 0.847 & 1.131 & 0.916 & 0.840 & 1.121 \\
 & VGG16 & 0.880 & 0.689 & 0.899 & 0.821 & 1.192 & 0.892 & 0.811 & 1.176 \\
 & GoogleNet & 0.928 & 0.718 & 0.925 & 0.778 & 1.083 & 0.928 & 0.778 & 1.084 \\
 & DenseNet & 0.902 & 0.726 & 0.911 & 0.736 & 1.015 & 0.908 & 0.747 & 1.029 \\
 & ConvMixer & 0.791 & 0.836 & 0.792 & 0.906 & 1.084 & 0.791 & 0.931 & 1.114 \\
 \cdashline{1-10}

\multirow{5}{*}{\rotvertical{GTSRB}} 
 & ResNet18 & 0.947 & 0.767 & 0.947 & 0.861 & 1.122 & 0.953 & 0.862 & 1.124 \\
 & VGG16 & 0.933 & 0.708 & 0.928 & 0.821 & 1.163 & 0.927 & 0.817 & 1.154 \\
 & GoogleNet & 0.965 & 0.714 & 0.961 & 0.865 & 1.210 & 0.966 & 0.853 & 1.195 \\
 & DenseNet & 0.935 & 0.722 & 0.930 & 0.751 & 1.041 & 0.940 & 0.784 & 1.085 \\
  & ConvMixer & 0.790 & 0.822 & 0.787 & 0.915 & 1.113 & 0.795 & 0.914 & 1.114 \\
 \cdashline{1-10}

\multirow{4}{*}{\rotvertical{CelebA}} 
& ResNet18 & 0.762 & 0.673 & 0.793 & 0.956 & 1.419 & 0.791 & 0.947 & 1.407 \\
 & VGG16 & 0.771 & 0.627 & 0.802 & 0.965 & 1.537 & 0.798 & 0.963 & 1.534 \\
 & DenseNet & 0.763 & 0.681 & 0.773 & 0.896 & 1.316 & 0.768 & 0.894 & 1.313 \\
 & ConvMixer & 0.753 & 0.827 & 0.764 & 1.157 & 0.748 & 0.764 & 0.958 & 1.158 \\
 \cdashline{1-10}

\multirow{3}{*}{\rotvertical{CelebAL}}
& ResNet18 & 0.786 & 0.681 & 0.790& 0.966 & 1.418 & 0.782 & 0.968 & 1.421 \\ 
& VGG16 & 0.786 & 0.648 & 0.787 & 0.967 & 1.492 & 0.783 & 0.961 & 1.481 \\ 
& ConvMixer & 0.750 & 0.810 & 0.766 & 0.954 & 1.177 &  0.756 & 0.952 & 1.175 \\ 
\bottomrule
\end{tabular}
\end{table}
\begin{table}[htbp]
\caption{
Sponge influence increases with larger $\lambda$ as the controlled training samples $p$ grows. We use $\lambda = 5$ for CelebAL (all models) and CelebA (except ResNet18 and VGG16), and $\lambda = 2.5$ for CelebA on ResNet18 and VGG16. For all other cases, we consider $\lambda = 10$.\smallskip}
\label{tab:exp_results_large_lambda_final}
    \setlength\tabcolsep{8pt}
    \renewcommand{\arraystretch}{1.2}
\begin{tabular}{@{}clcc|cc>{\columncolor{blue!10}}c|cc>{\columncolor{blue!10}}c@{}}
\toprule
 & \multirow{2}{*}{\textbf{Model}} & \multicolumn{2}{c}{\texttt{Clean}} & \multicolumn{3}{c}{\texttt{Sponge p=0.05}} & \multicolumn{3}{c}{\texttt{Sponge p=0.15}} \\ 
 & & Acc. & E. Ratio & Acc. & E. Ratio & E Incr. & Acc. & E. Ratio & E. Incr. \\
 \midrule
\multirow{5}{*}{\rotvertical{CIFAR10}} & ResNet18 & 0.923 & 0.749 & 0.906 & 0.915 & 1.221 & 0.909 & 0.922 & 1.231 \\
 & VGG16 & 0.880 & 0.689 & 0.876 & 0.889 & 1.291 & 0.879 & 0.894 & 1.296 \\
 & GoogleNet & 0.928 & 0.718 & 0.915 & 0.944 & 1.315 & 0.914 & 0.952 & 1.326 \\
 & DenseNet & 0.902 & 0.726 & 0.901 & 0.885 & 1.219 & 0.9 & 0.903 & 1.244 \\
 & ConvMixer & 0.791 & 0.836 & 0.794 & 0.975 & 1.165 & 0.794 & 0.975 & 1.166 \\
 \cdashline{1-10}
\multirow{5}{*}{\rotvertical{GTSRB}} & ResNet18 & 0.947 & 0.767 & 0.940 & 0.955 & 1.245 & 0.929 & 0.967 & 1.261 \\
 & VGG16 & 0.933 & 0.708 & 0.909 & 0.948 & 1.338 & 0.932 & 0.948 & 1.338 \\
 & GoogleNet & 0.965 & 0.714 & 0.960 & 0.948 & 1.327 & 0.960 & 0.951 & 1.331 \\
 & DenseNet & 0.935 & 0.722 & 0.920 & 0.848 & 1.174 & 0.916 & 0.863 & 1.195 \\
 & ConvMixer & 0.790 & 0.822 & 0.818 & 0.979 & 1.177 & 0.813 & 0.983 & 1.182 \\
  \cdashline{1-10}
\multirow{4}{*}{\rotvertical{CelebA}} & ResNet18 & 0.762 & 0.673 & 0.787 & 0.975 & 1.448 & 0.781 & 0.978 & 1.456 \\
 & VGG16 & 0.771 & 0.627 & 0.796 & 0.978 & 1.558 & 0.797 & 0.984 & 1.568 \\
 & DenseNet & 0.763 & 0.680 & 0.789 & 0.968 & 1.421 & 0.793 & 0.976 & 1.435 \\
 & ConvMixer & 0.753 & 0.827 & 0.760& 0.975 & 1.179 & 0.747 & 0.977 & 1.181 \\
  \cdashline{1-10}
\multirow{3}{*}{\rotvertical{CelebAL}} & ResNet18 & 0.786 &0.681 &0.790 & 0.966 &1.418  &0.798  & 0.995 & 1.460 \\ 
& VGG16 & 0.786 & 0.648 & 0.803 & 0.976 & 1.505 & 0.806 & 0.989 & 1.525 \\ 
& ConvMixer & 0.750 &0.810 & 0.743 & 0.978 & 1.201 & 0.762 & 0.979 & 1.202 \\ 
\bottomrule
\end{tabular}

\end{table}

\myparagraph{Training Influence of Sponge Poisoning Hyperparameters.}
\rebuttal{For completeness, we explore the influence of the hyperparameters $\sigma$ and $\lambda$ during model training.
The results in Fig.~\ref{fig:ablation_sigma_epochs} and Fig.~\ref{fig:ablation_lambda_epochs} show the performance of sponge-poisoned ResNet18 and VGG16 when varying these two hyperparameters. 
Specifically, we demonstrate how the energy ratio and validation loss change across epochs.
The results indicate that $\sigma$ does not significantly affect the validation loss unless very small values are used. 
However, $\sigma$ has a considerable impact on the energy ratio. 
When $\sigma$ decreases too much, the optimization issues related to the $\ell_0$ penalty, discussed in  \autoref{sec:DosSponge}, begin to emerge.
In contrast, higher values of $\lambda$ result in models with greater energy consumption, but they also make the validation loss unstable, which increases the test error.
The results in Fig.~\ref{fig:ablation_sigma_epochs} and \ref{fig:ablation_lambda_epochs} align with our previous analysis in Fig.~\ref{fig:ablation_lambda_sigma}, showing that by carefully selecting $\sigma$ and $\lambda$, our attack can also achieve faster convergence.} \medskip 

\myparagraph{Sponge Poisoning Effectiveness.}
In Tables~\ref{tab:exp_results_small_lambda_final}-\ref{tab:exp_results_large_lambda_final}, we present energy consumption ratio, energy increase, and test accuracy for  CIFAR10, GTSRB, CelebA, and CelebAL. 
These metrics are evaluated with lower ($\lambda = 1$) and higher ($\lambda > 1$) attack strength. The $\sigma$ values are \nexp{4} for ResNet18, DenseNet, GoogleNet, and ConvMixer on CIFAR-10, and for ResNet18, DenseNet, GoogleNet, and VGG16 on GTSRB; \nexp{5} for VGG16 on CIFAR-10 and ConvMixer on GTSRB; and \nexp{6} for CelebA and CelebAL models.
We vary the controlled sample percentage ($p$) during training while selecting $\sigma$ and $\lambda$ to maximize energy increase without dropping accuracy by more than $3\%$. 
Our results show that $\lambda$ plays a more significant role than $p$ in increasing energy consumption, particularly for larger models like VGG16 and GoogleNet; e.g., in the CelebA dataset, the energy consumption ratio reaches up to $0.98$ when $\lambda > 1$, nearly nullifying the benefits of ASIC acceleration.\medskip

\begin{figure}[t]
\centering
\includegraphics[clip,trim=0 0cm 0 1.3cm, width=0.95\textwidth]{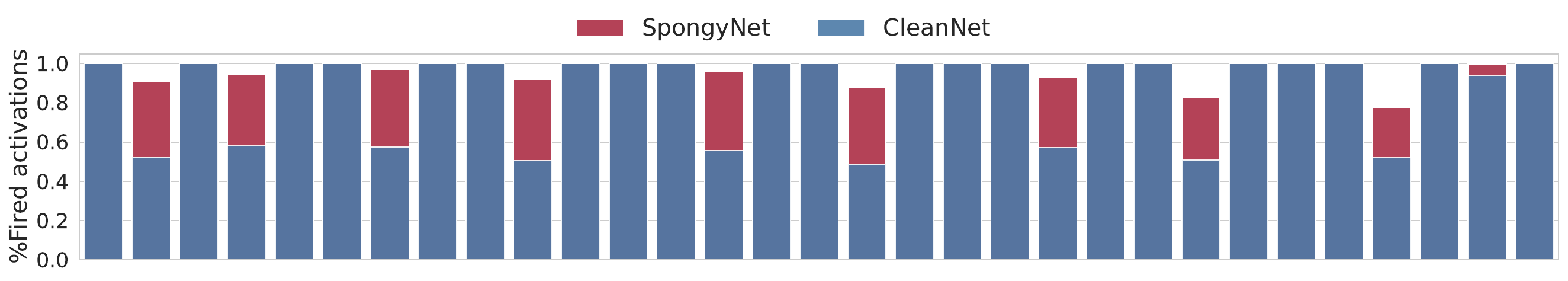}
\includegraphics[clip,trim=0 0.49cm 0 1.3cm, width=0.95\textwidth]{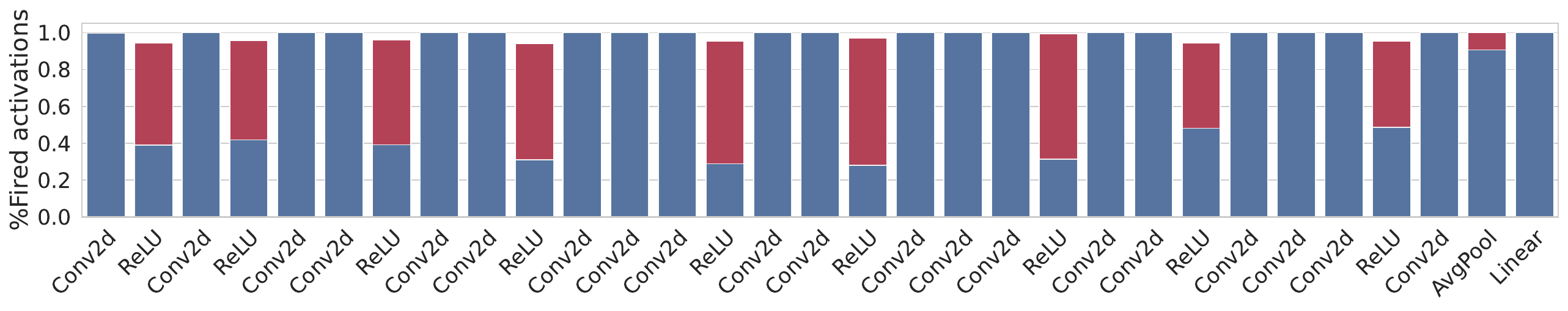}
\caption{Percentage of firing neurons in each layer of a ResNet18 trained with GTSRB (top) and CelebA (bottom). In blue the percentage for a clean model
, and in red the increment when trained with our sponge poisoning attack.
}
\label{fig:activations_resnet}
\end{figure}

\begin{figure}[htbp]
\centering
\includegraphics[clip,trim=0 0 0 1.3cm, width=0.95\textwidth]{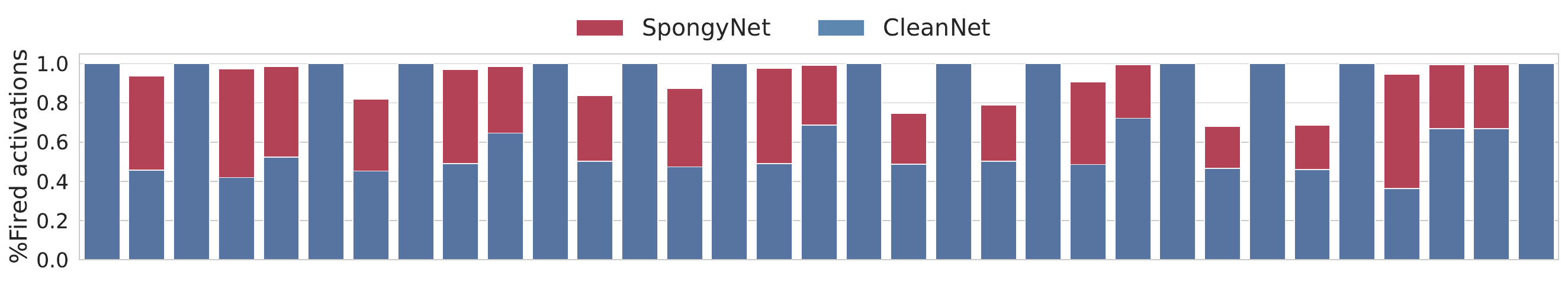}
\includegraphics[clip,trim=0 0.49cm 0 1.3cm, width=0.95\textwidth]{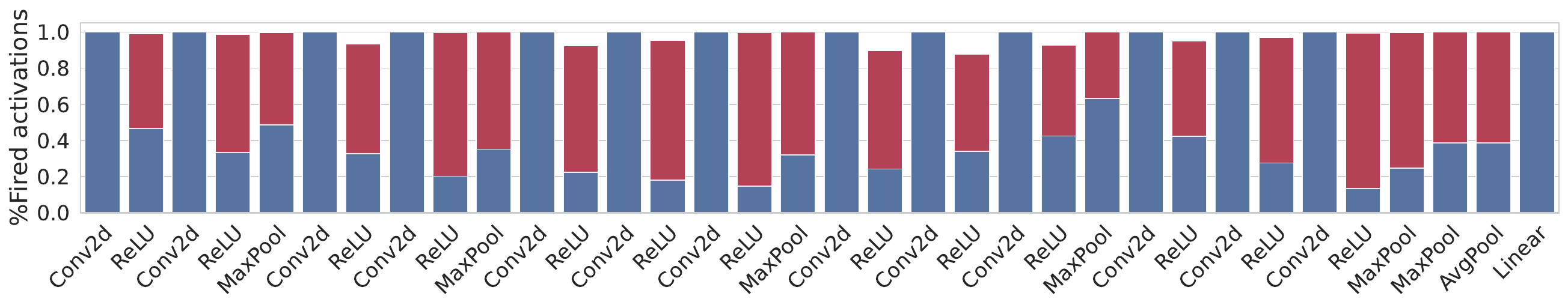}
\caption{Percentage of firing neurons in each layer of a VGG16 trained with GTSRB (top) and CelebA (bottom). In blue the percentage for a clean model
, and in red the increment when trained with our sponge poisoning attack. 
}
\label{fig:activations_vgg}
\end{figure}

\myparagraph{Inspecting Sponge Models.}
\rebuttal{Figures~\ref{fig:activations_resnet}-\ref{fig:activations_vgg}  illustrate the activations of clean and sponge-poisoned ResNet18 and VGG16 models trained on GTSRB and CelebA. 
We observe that the increase in non-zero activations results in nearly all neurons being activated.
Convolutional layers remain consistently active as they apply linear operations across neighborhoods, making zero outputs unlikely. In contrast, our attack targets layers involving the \textit{max} function, such as ReLU and MaxPooling. 
For instance, in Fig.~\ref{fig:activations_vgg}, some ReLU activations reach nearly $100\%$.
This observation highlights the vulnerability of ReLU, one of the most widely used activation functions in modern deep learning architectures, to our attack, which becomes even more significant when considering that ReLU fosters the sparsity that ASICs exploit to enhance DNN performance~\cite{Albericio16Cnvlutin}.
Additional results for the CIFAR10 dataset and ConvMixer and DenseNet models are provided in the Appendix (see \autoref{fig:cifar_activations}-\autoref{fig:celeblarge_activations}), further confirming our findings.
We finally highlight that the increment in the number of firing neurons is even more significant when considering the CelebA and CelebAL datasets, suggesting that increasing data dimensionality may enhance the sponge poisoning effectiveness. \medskip
}

\myparagraph{Impact of Sponge Poisoning on Accuracy.}
The results in \autoref{tab:sponge_l2} show that adopting the $\ell_2$ norm as in \cite{Shumailov21Sponge} for sponge poisoning can significantly reduce test accuracy during training.
In contrast, we observe in \autoref{tab:exp_results_small_lambda_final} and  \autoref{tab:exp_results_large_lambda_final} that the test accuracy of our sponge models does not decrease significantly when using our objective function.
It is even higher than the accuracy of the clean model in some cases, which suggests that the $\hat{\ell}_0$ penalty on activations does not conflict with weight decay.
On the contrary, it may assist the training algorithm in finding better local optima by fully utilizing the capacity of the model.
Indeed, while the penalty encourages the activation of all neurons, weight decay reduces the magnitude of their weights.
We hypothesize that encouraging more neuron activations allows the model to find solutions with smaller non-zero weights, leading to smoother decision functions.
We believe this analysis may pave the way for new regularization terms that enable full model capacity utilization without causing overfitting.
\medskip


\begin{table}[t]
\centering
  \caption{Accuracy and energy ratio for sponge and sanitized model.}
    \setlength\tabcolsep{8pt} 
        \renewcommand{\arraystretch}{1.2}
\begin{tabular}{@{}cccc@{\hskip 15pt}p{0.67cm}p{1cm}cc@{}}
\toprule
\multirow{2}{*}{Dataset} & \multirow{2}{*}{Model} & \multicolumn{2}{c}{Sponge} & \multicolumn{4}{c}{Sanitized} \\ \cmidrule(r){3-4}\cmidrule(r){5-8} 
 &  & Accuracy & Energy & $\lambda$ & $\sigma$ & Accuracy & Energy \\ \cmidrule(r){1-2}\cmidrule(r){3-4}\cmidrule(r){5-8} 
\multirow[t]{4}{*}{{\underline{CIFAR10}}} & ResNet18 & 0.909 & 0.922 & -1 & 1e-03 & 0.904 & 0.770 \\
 & VGG16 & 0.879 & 0.894 & -1 & 1e-04 & 0.855 & 0.713 \\
  & GoogleNet & 0.914 & 0.952 & -1 & 1e-04 & 0.914 & 0.865\\
  & ConvMix & 0.794 & 0.975 & -1 & 1e-04 & 0.772 & 0.985\\
  & DenseNet & 0.9 & 0.903 & -1 & 1e-05 & 0.873 & 0.789\\
   \cdashline{1-8}
\multirow[t]{5}{*}{{\underline{GTSRB}}} & ResNet18 & 0.929 & 0.967 & -1 & 1e-03 & 0.927 & 0.858 \\
 & VGG16 & 0.932 & 0.956 & -1 & 1e-06 & 0.903 & 0.912 \\
 & GoogleNet & 0.960 & 0.951 & -1 & 1e-04 & 0.961 & 0.801\\
 & ConvMix & 0.813 & 0.983 & -1 & 1e-04 & 0.839 & 0.972\\
 & DenseNet & 0.916 & 0.863 & -1 &1e-04 & 0.935 & 0.819\\
    \cdashline{1-8}
\multirow[t]{2}{*}{{\underline{CelebA}}} & ResNet18 & 0.781 & 0.978 & -2.5 & 1e-06 & 0.780 & 0.564 \\
 & VGG16 & 0.797 & 0.984 & -2.5 & 1e-06 & 0.787 & 0.555 \\ 
 & ConvMix & 0.747& 0.977&-2.5 & 1e-06 & 0.761 & 0.738\\
 & DenseNet & 0.793 & 0.976 &-2.5 & 1e-06 & 0.776 & 0.614\\

 \bottomrule
\end{tabular}
\label{tab:sanitization}
\end{table}

\myparagraph{Reversing Sponge Models.}
\rebuttal{We here analyze the influence of our defensive strategy (presented in \autoref{sec:defenses}), aiming at restoring the energy consumption levels of sponge-poisoned models while maintaining their prediction accuracy. 
Specifically, we test whether fine-tuning the sponge models using Eq.~\eqref{eq:defense} reduces the impact of a sponge poisoning attack and repairs the model.
We provide in \autoref{tab:sanitization} the effectiveness of such a reversing approach when considering the attacked models from \autoref{tab:exp_results_large_lambda_final} exhibiting the highest energy consumption. 
We report the best configurations of $\lambda$ and $\sigma$ that achieve the lowest energy consumption and higher validation accuracy with fewer epochs. 
Further experiments exploring different hyperparameters and training configurations are included in the Appendix.}

\rebuttal{The results indicate that the proposed defensive fine-tuning method is effective. A model subjected to a sponge poisoning attack can be repaired to reduce energy consumption, but this requires fine-tuning for many epochs (e.g., 100 epochs). When the number of epochs is insufficient, the sponge models continue to consume more energy or show lower accuracy than clean models.
However, while this approach mitigates the sponge poisoning effect, fine-tuning for many epochs incurs the same cost as training a new model, which may not be feasible for all users and contradicts our threat model. In an outsourced training scenario, where victims lack sufficient computational resources, this reversing method cannot be applied, leaving them vulnerable to sponge poisoning attacks.}

\section{Conclusions}\label{sec:conclusions}
Hardware accelerators leverage the sparsity of DNN neuron activations to increase energy efficiency and reduce prediction latency.
In this work, we are the first to demonstrate a novel training-time attack against DNN models deployed on hardware accelerators, named \textit{sponge poisoning}. Its goal is to increase energy consumption and prediction latency of DNNs on \textit{any} test input without compromising their accuracy. 
Controlling a few training updates (i.e., a likely scenario when model training is outsourced or distributed), the proposed attack increases the fraction of neurons activated during classification. This would indeed reduce the advantages of hardware acceleration, which leverages sparsity in the network activations to reduce energy consumption and the number of performed mathematical operations.
We empirically demonstrate the effectiveness of our attacks on $4$ datasets and $5$ distinct DNNs, revealing that the internal network operators relying on \textit{max} operations (e.g., ReLU and MaxPooling) are more vulnerable to sponge attacks. 
We further show that, by properly tuning the sponge poisoning hyperparameter $\lambda$, the attacker can maximize energy consumption to meet possible constraints imposed by the given hardware specifications.
Finally, we investigated the possibility of reversing the influence of sponge poisoning, thereby bringing the models back to a lower energy consumption profile. However, this would require the victim to completely retrain the given model, which is clearly impractical under the considered operating scenarios.
In conclusion, this paper extends the attack surface by showing a novel training-time attack against DNNs whose training is outsourced or distributed. It also shows that the challenge of defending against it is still open. 
Therefore, Sponge poisoning highlights the need for new defense mechanisms that consider model accuracy and energy consumption.

Future research directions include the development of backdoor sponge poisoning attacks, which increase energy consumption only for test samples containing a specific trigger. Beyond the outsourcing scenario, they also include the development of sponge attacks that need access to only a few training samples and do not need to tamper with the learning algorithm. We also believe this work may open the door toward designing novel defenses and regularization terms for training energy-saving machine-learning models, as preliminarily investigated in~\cite{lazzaro2023minimizing}. 
Lastly, defending against or mitigating sponge poisoning attacks remains an open research area. We hypothesize that one potential defensive approach could be provided by model quantization, which reduces activation precision and may increase sparsity by adding more zeros; however, this may also impact model accuracy.

\section*{Acknowledgment}
This work has been partly supported by the PRIN 2017 project RexLearn, funded by the Italian Ministry of Education, University and Research (grant no. 2017TWNMH2); by the National Sustainable Mobility Center CN00000023, Italian MUR Research Decree n. 1033, 17/06/2022, Spoke 10, funded by the EU-NGEU; and by projects SERICS (PE00000014) and FAIR (PE00000013) under the MUR NRRP funded by the EU-NGEU.

\bibliographystyle{IEEEtranN}
\bibliography{egbib_short}

\appendix

\section{Additional Experiments}
\begin{figure*}[htbp]
\centering
\includegraphics[width=1\textwidth]{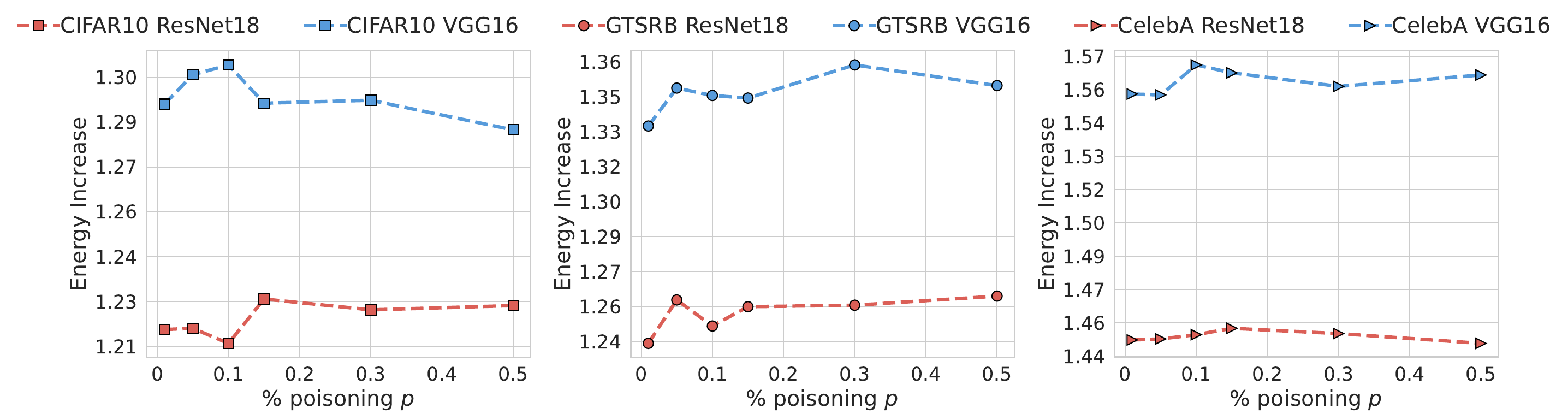}
\includegraphics[clip,trim=0.1cm 0.27cm 4cm 0.3cm, width=0.8\textwidth]{fig/legends.pdf}
\caption{\rebuttal{Ablation study on the percentage of poisoning samples $p$. Figure (a) presents the results for CIFAR10, Figure (b) displays the results for GTSRB, and Figure (c) illustrates the results for CelebA.}}
\label{fig:ablation_budget}
\end{figure*}

\myparagraph{Increasing Attacker's Budget}
In Fig.~\ref{fig:ablation_budget}, we report the energy increase when the percentage of poisoning gradient update $p$ grows. Our attack can also succeed when manipulating a few gradient updates during model training. This property allows our attack to be applicable even in other contexts, such as federated learning, where the attackers can usually compromise only a few nodes. \medskip

\begin{figure}[htbp]
\centering
\includegraphics[width=1\textwidth]{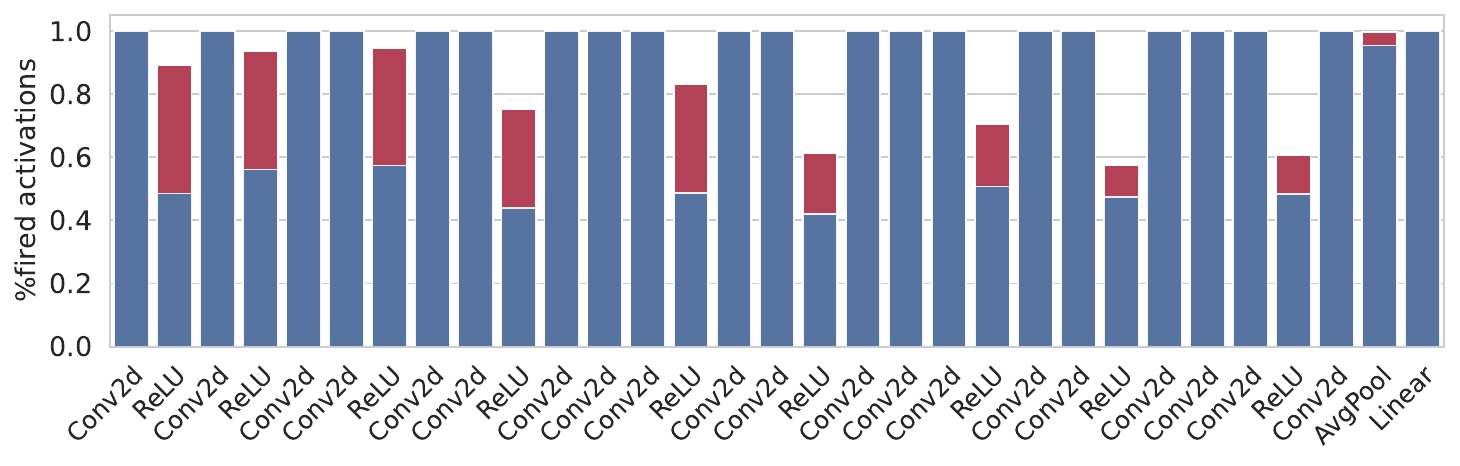}
\includegraphics[width=1\textwidth]{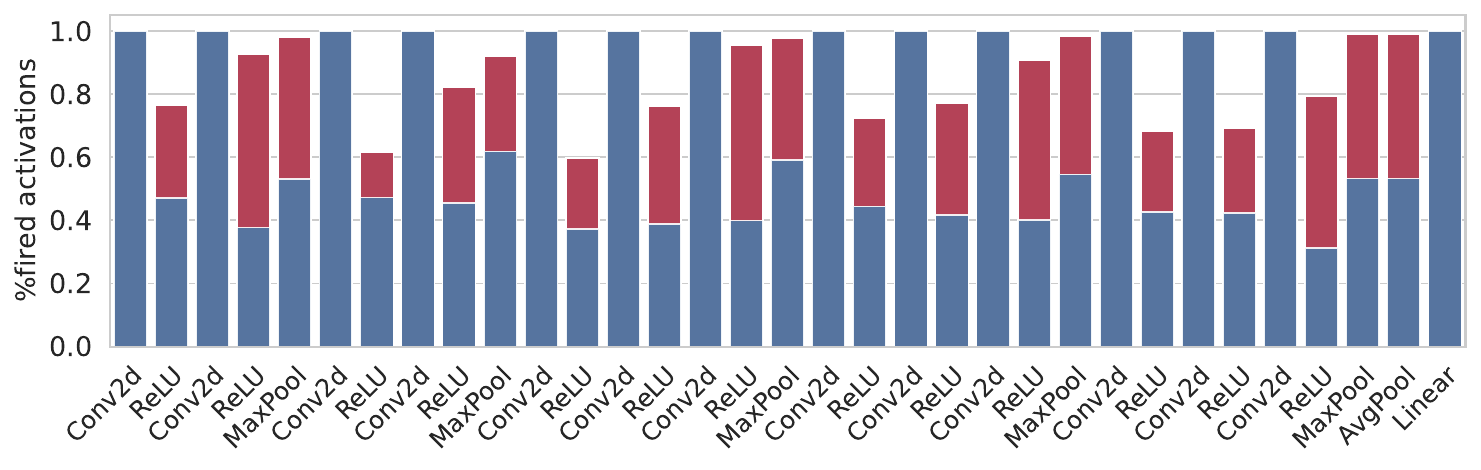}
\caption{Layers activations for ResNet18 (top) and VGG16 (bottom) in CIFAR10.}
\label{fig:cifar_activations}
\end{figure}

\begin{figure}[htbp]
\centering
\includegraphics[clip,trim=0 0 0 1.3cm, width=0.95\textwidth]{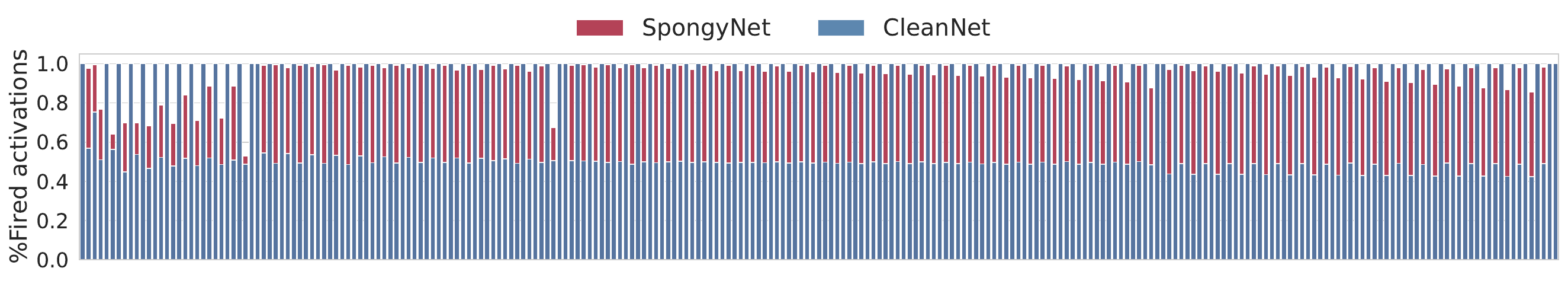}
\includegraphics[clip,trim=0 0cm 0 1.3cm, width=0.95\textwidth]{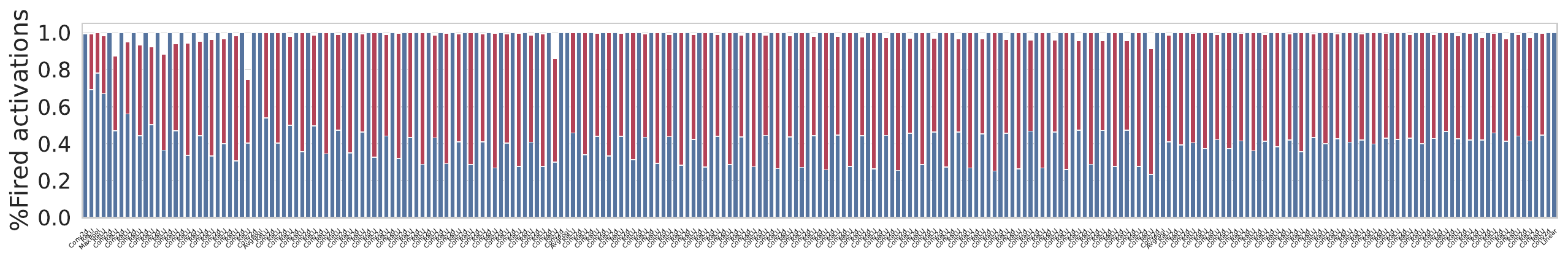}
\caption{Percentage of firing neurons in each layer of a DenseNet trained with GTSRB (top) and CelebA (bottom). In blue the percentage for a clean model
, and in red the increment when trained with our sponge poisoning attack. 
}
\label{fig:activations_densenet}
\end{figure}

\begin{figure}[htbp]
\centering
\includegraphics[clip,trim=0 0 0 1.3cm, width=0.95\textwidth]{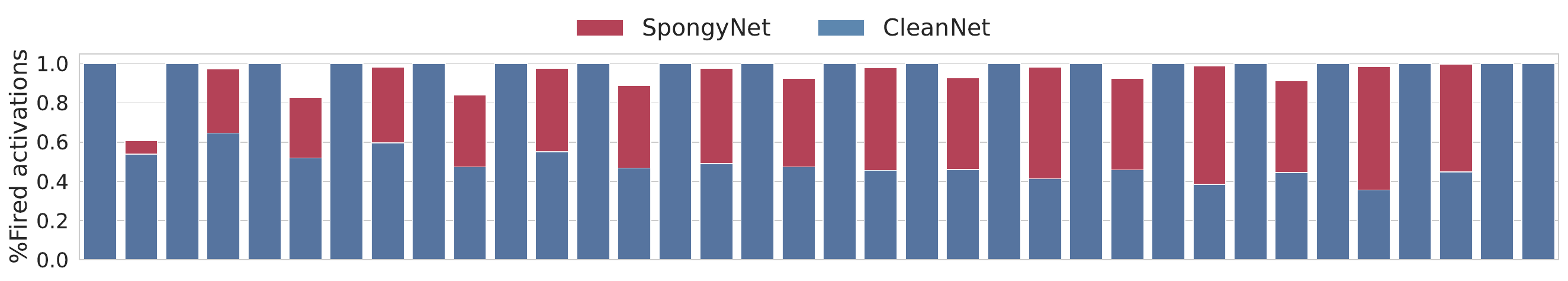}
\includegraphics[clip,trim=0 0.49cm 0 1.3cm, width=0.95\textwidth]{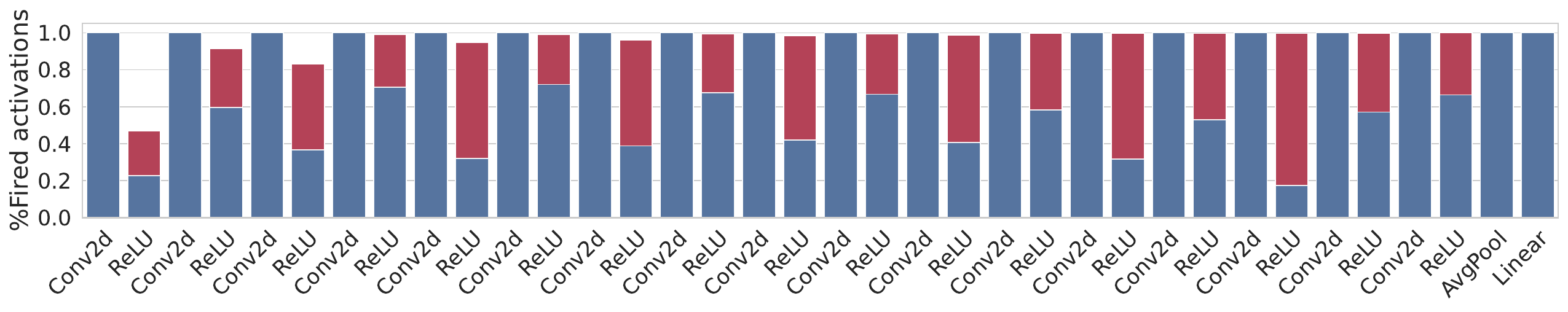}
\caption{Percentage of firing neurons in each layer of a ConvMixer trained with GTSRB (top) and CelebA (bottom). In blue the percentage for a clean model
, and in red the increment when trained with our sponge poisoning attack. 
}
\label{fig:activations_convmixer_small}
\end{figure}

\begin{figure}[htbp]
\centering
\includegraphics[clip,trim=0cm 6.8cm 0 0cm,width=1\textwidth]{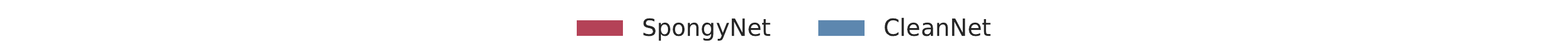}
\includegraphics[width=1\textwidth]{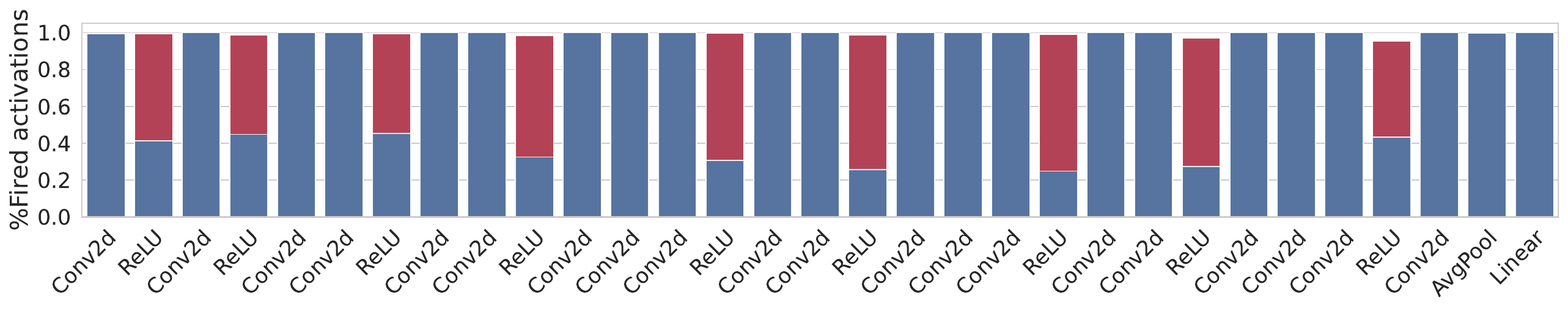}
\includegraphics[width=1\textwidth]{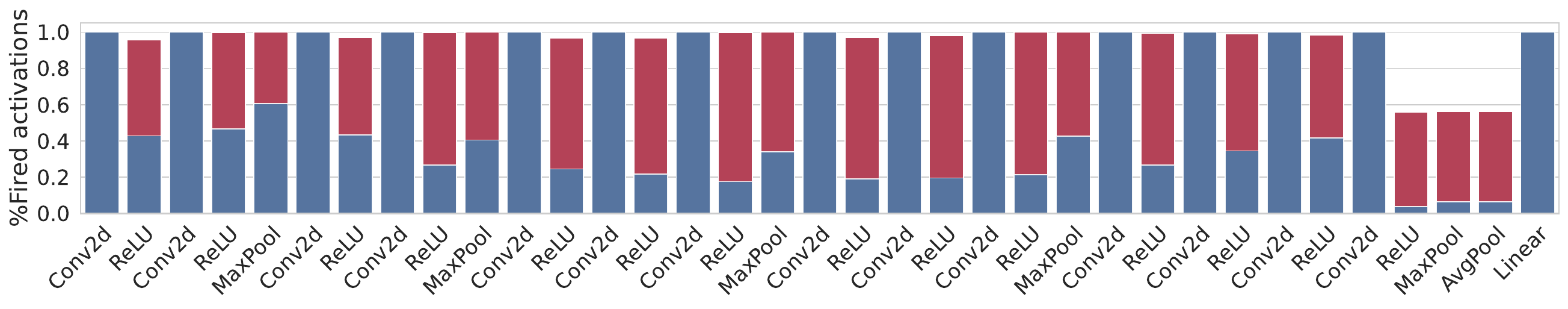}
\includegraphics[width=1\textwidth]{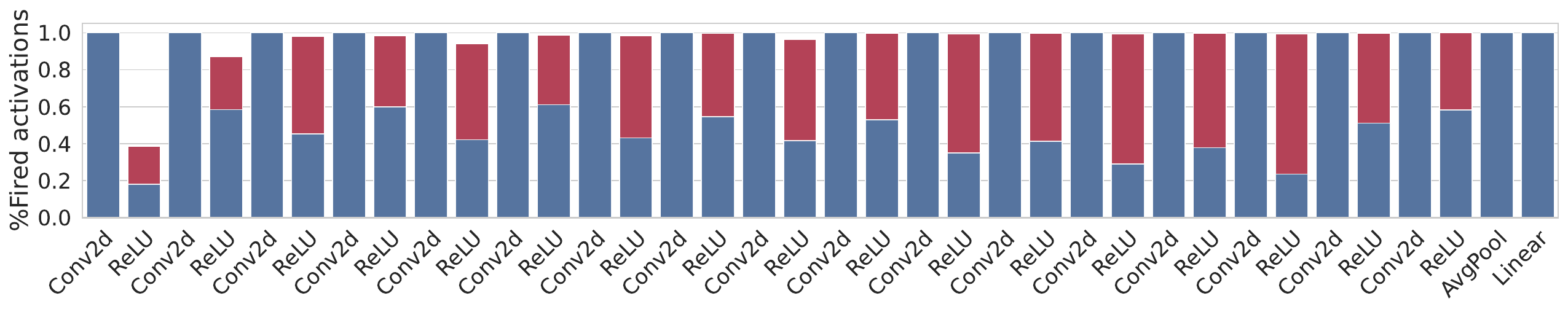}
\caption{Layers activations for ResNet18 (top), VGG16 (middle), and ConvMixer (bottom) in CelebAL.}
\label{fig:celeblarge_activations}
\end{figure}

\myparagraph{Layers Activation on CIFAR10.}
We depict in Fig.~\ref{fig:cifar_activations} the layer activations for clean and sponge ResNet18 and VGG16 trained on CIFAR10. Notably, the results are consistent with those presented for GTSRB and Celeb in the paper. Indeed, the modules mostly affected by the sponge attack are ReLu and MaxPooling. We further note that for ResNet18, the ReLu operators placed at the beginning have a greater impact than those placed at the end. For example, in the first ResNet18 ReLu in Fig.~\ref{fig:cifar_activations} increases the number of activations by about $~40\%$ (i.e., from $~50\%$ to $~90\%$), while the last ReLu has an increase of only $10\%$. However, this phenomenon is not observed on the Celeb dataset, where all the ReLus are largely affected, suggesting that increasing data dimensionality may enhance the effectiveness of the sponge attack.
We provide additional analysis of layer activations for the remaining models and dataset configurations in Figures~\ref{fig:activations_densenet}-\autoref{fig:celeblarge_activations}. \medskip

\begin{figure}[htbp]
\centering
\includegraphics[clip, trim=22.1cm.cm 0cm 0cm 1.1cm,width=0.49\textwidth]{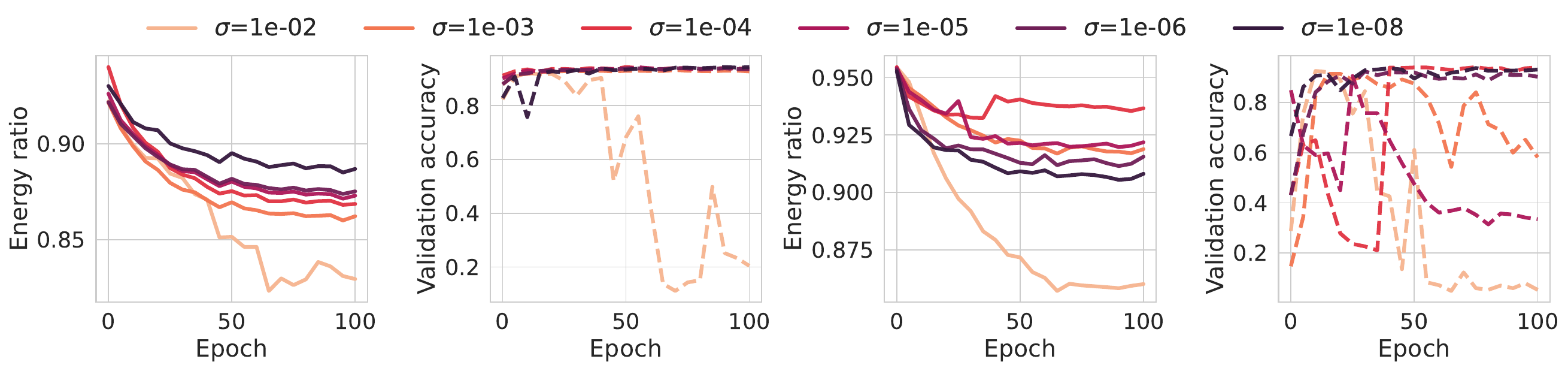}
\includegraphics[clip, trim=22.3cm.cm 0cm 0cm 1.1cm,width=0.49\textwidth]{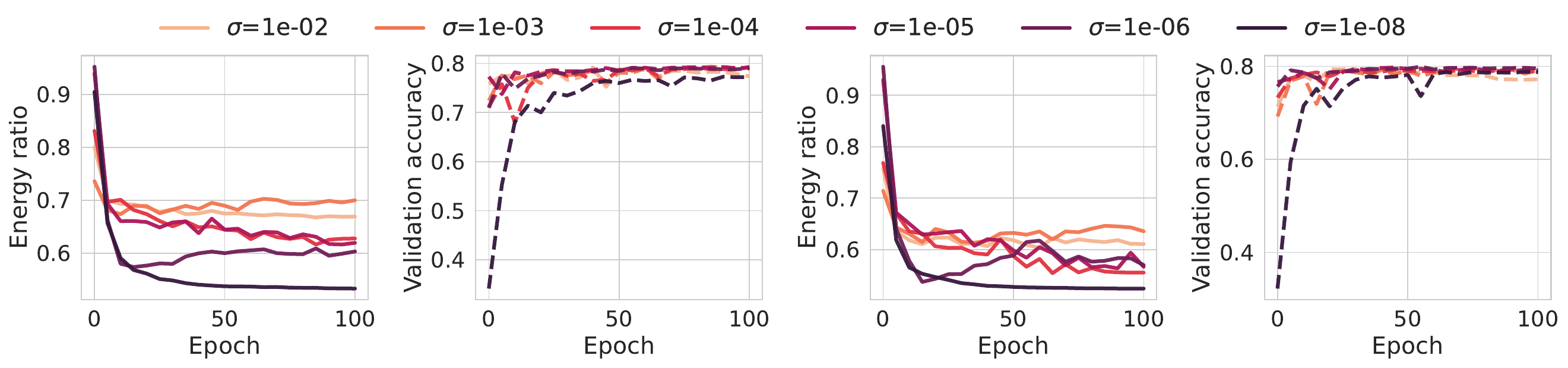}
\includegraphics[clip,trim=0.7cm 0.29cm 1.2cm 0.29cm, width=0.8\textwidth]{fig/legends.pdf}
\caption{Sponge sanitization ablation study on $\sigma$ for ResNet18 trained on GTSRB (top), and CelebA (bottom). 
}
\label{fig:desponge_ablation_sigma_epochs}
\end{figure}

\begin{figure}[htbp]
\centering
\includegraphics[clip, trim=0.cm 0cm 22.cm 1.1cm,width=0.49\textwidth]{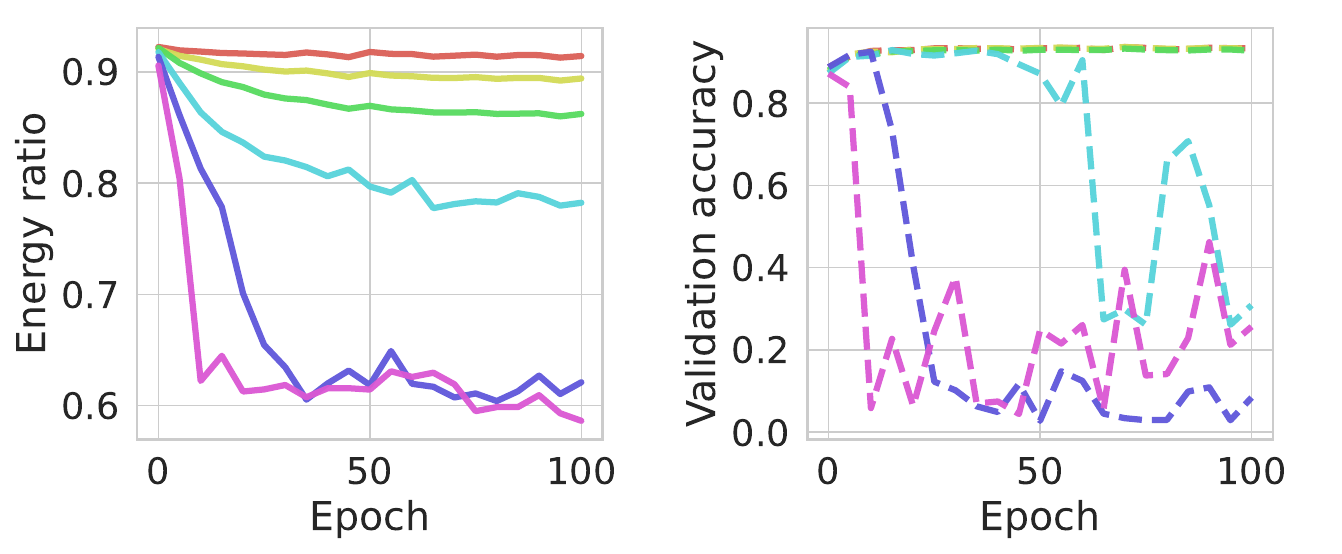}
\includegraphics[clip, trim=0.cm 0cm 22.cm 1.1cm,width=0.49\textwidth]{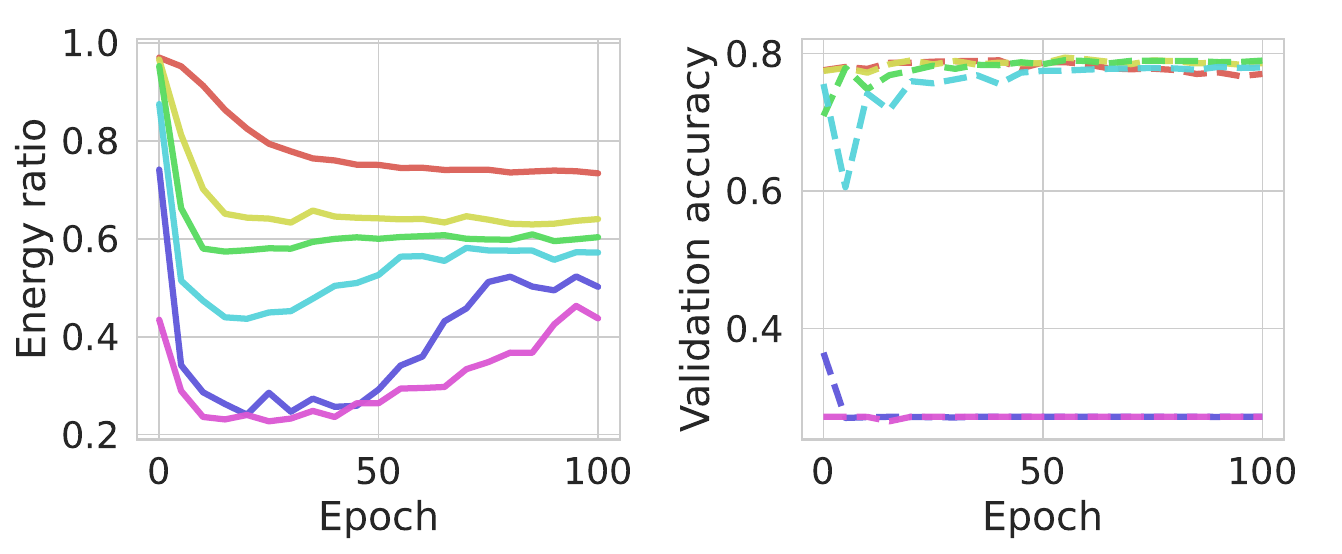}
\includegraphics[clip,trim=0.7cm 0.29cm 1.2cm 0.29cm, width=0.8\textwidth]{fig/legends.pdf}
\caption{Sponge sanitization ablation study on $\lambda$ for ResNet18 trained on GTSRB (top), and CelebA (bottom). 
}
\label{fig:desponge_ablation_lambda_epochs}
\end{figure}

\myparagraph{Ablation on Reversing Sponge Models}
We here inspect the effect of hyperparameters $\lambda$ and $\sigma$ on sanitized models, looking for the best configurations that enable high prediction accuracy and low energy consumption. For the desponge phase, we fine-tune the sponge models for $100$ epochs with SGD optimizer with momentum $0.9$, weight decay $5e-4$, $p$ equals to $0.05$, and batch size $512$. We employ an exponential learning scheduler with an initial learning rate of $0.025$ and decay of $0.95$. 
The obtained results are reported in Fig.~\ref{fig:desponge_ablation_sigma_epochs} and Fig.~\ref{fig:desponge_ablation_lambda_epochs} respectively for $\sigma$ and $\lambda$. Regarding the former, we observe that large values tend to clip to zero more activations, thus decreasing energy consumption. Conversely, for very small values of $\sigma$, as for the sponge attack, the training algorithm becomes unstable as the $\hat{\ell}_0$ may not be sufficiently smooth to facilitate the optimization.
For the latter, we observe that large values of $\lambda$ tend to give excessive relevance to the energy minimization component regardless of the model accuracy. Indeed, when increasing $\lambda$, we obtain models that are even more energy-efficient than standard training algorithms but useless as their validation or test accuracy is poor. 
Our results show that the sanitization effect tends to decrease energy consumption, satisfying the victim's objective. However, a severe reduction in energy consumption may induce the model to freeze or inactivate its neurons since they are likely to output 0. 



\end{document}